\documentclass[12pt,pre,groupedaddress,showpacs]{revtex4}

\usepackage{amsmath}
\usepackage{graphicx}
\usepackage{verbatim}

\newcommand{\mf}{mean-field}
\newcommand{\nmf}{near-mean-field}
\newcommand{\nnmf}{(near-)mean-field}
\newcommand{\e}{\epsilon}
\newcommand{\es}{\varepsilon} 
\newcommand{\dhs}{\Delta h}
\renewcommand{\lg}{Landau-Ginzburg}
\newcommand{\lgw}{Landau-Ginz\-burg-Wilson}
\newcommand{\ps}{Parisi-Sourlas}

\renewcommand{\t}{\tilde}
\newcommand{\p}{\phi}
\newcommand{\chit}{\chi}

\newcommand{\pfc}{\phi_{\rm fc}}
\newcommand{\fluc}{\phi_{\rm f}}
\newcommand{\tp}{\tilde\phi}
\newcommand{\gp}{R^{d}\e^{2-d/2}}
\newcommand{\gps}{R^{d}\Delta h^{3/2 - d/4}}
\newcommand{\tk}{\tilde k}
\newcommand{\pb}{P_B} 
\newcommand{\half}{\frac12}
\newcommand{\pa}{\partial}
\newcommand{\pxt}{\psi_F(\xv,t)} 
\newcommand{\pbxt}{{\bar \psi}_F(\xv,t)}
\newcommand{\bpsi}{{\bar \psi}_F}
\newcommand{\phxt}{\p(\xv,t)}
\newcommand{\grass}{\psi_F}
\newcommand{\grassbar}{{\bar \psi}_F}

\newcommand{\chc}{Cahn-Hilliard-Cook}
\renewcommand{\sp}{spin\-odal}
\newcommand{\psp}{pseudo\-spin\-odal}
\newcommand{\phix}{\phi(\xv)}
\newcommand{\bsig}{ \psi}
\newcommand{\orders}{\psi}
\newcommand{\gpst}{R^{d}\es^{3 - d/2}}

\newcommand{\lb}{{<}}
\newcommand{\rb}{{>}}

\newcommand{\hp}{h_{\rm P}}
\newcommand{\fc}{fun\-da\-mental clus\-ter}
\newcommand{\xv}{\mathbf{x}}
\newcommand{\yv}{\mathbf{y}}
\newcommand{\kv}{\mathbf{k}}
\newcommand{\kvt}{\tilde{\mathbf{k}}}
\newcommand{\tvk}{\kvt}
\newcommand{\tfc}{\tau_{\rm f, c}} 
\newcommand{\tfs}{\tau_{\rm f, s}} 
\newcommand{\tobjc}{\tau_{\rm fc, c}} 
\newcommand{\tobjs}{\tau_{\rm fc, s}} 
\newcommand{\jp}{J_{\rm P}}

\begin{document}

\title{The structure of fluctuations near mean-field critical points and
spinodals and its implication for physical processes}

\author{W.\ Klein}
\affiliation{Department of Physics and Center for
Computational Science, Boston University, Boston, Massachusetts 02215, USA}

\author{Harvey Gould}
\affiliation{Department of Physics, Clark University, Worcester, Massachusetts
01610, USA}

\author{Natali Gulbahce}
\affiliation{Theoretical Division and Center for
Nonlinear Studies, Los Alamos National Laboratory, Los Alamos, New
Mexico 87545, USA}
\altaffiliation[Formerly at ]{Department of Physics, Clark University, Worcester, Massachusetts
01610, USA}

\author{J. B. Rundle}
\affiliation{Department of Physics and Center for Computational 
Science and Engineering, University of California, Davis, California 95616, USA}

\author{K. Tiampo}
\affiliation{Earth Science Department, University of Western Ontario,
London, Ontario N6A 5B7, Canada}

\begin{abstract}
We analyze the structure of fluctuations near critical points and
spinodals in
\mf\ and
\nmf\ systems. Unlike systems that are non-\mf, for which a fluctuation
can be represented by a single cluster in a properly chosen percolation
model, a fluctuation in \mf\ and \nmf\ systems consists of a large number
of clusters, which we term fundamental clusters. The structure of the
latter and the way that they form fluctuations has important
physical consequences for phenomena as diverse as nucleation in supercooled liquids, spinodal
decomposition and continuous ordering, and the statistical distribution of
earthquakes. The effects due to the 
fundamental clusters 
implies that they are physical objects and not only
mathematical constructs.
\end{abstract}

\pacs{64.70.Pf, 05.70.Jk, 64.60.Fr, 64.60.My}

\maketitle

\section{\label{sec1}Introduction}
Systems that are \mf\ or \nmf\ are common in nature. Examples of such systems include metals with
long-range elastic forces~\cite{shenoy,kleinetal1}, earthquake faults with
long-range stress transfer Green's functions~\cite{kleinetal2}, and
polymers~\cite{binder}. The connection between the range of the interaction
and \mf\ behavior was made by Kac and collaborators~\cite{kacetal} who noted
that a system with a pairwise additive potential of the form
\begin{equation}
\label{kacpot}
V(x) = V_{R}(x) +\gamma^{d}\Phi(\gamma x),
\end{equation}
becomes \mf\ in the limit $\gamma \rightarrow 0$. In Eq.~\eqref{kacpot} $x
= |\xv|$,
$V_{R}(x)$ is a short range reference potential,
and $d$ is
the spatial dimension. The limit 
$\gamma \rightarrow 0$ is taken after the thermodynamic limit and before a
critical point is approached. It is also required
that~\cite{kacetal}
\begin{equation}
\label{intkac}
\int\!d\xv\,\gamma^{d}|\Phi(\gamma x)| < \infty.
\end{equation}
so that the energy per particle or spin remains finite in the $\gamma\rightarrow 0$ limit.
The interaction range $R$ is defined by the second moment of the potential,
\begin{equation}
\label{range}
R^{2} \propto \!\int\!d \xv\, x^{2}\gamma^{d}\Phi(\gamma x) \propto
\gamma^{-2}.
\end{equation}
Hence, as $\gamma\rightarrow 0$, $R \to \infty$. We will refer to systems with $R \gg 1$ but
not infinite as {\it \nmf}; systems with $R\rightarrow \infty$
are
{\it \mf}~\cite{tsallis}.

The kinetics of phase transitions is different in systems with $R \gg 1$ than in systems with $R\sim
1$. For example, nucleation in the former often occurs near a
\psp~\cite{Heermann, Novotny, Gulbahce} where the
surface tension is small, which results in a nucleating droplet that has a
different structure~\cite{heermannkl, ungerkl, monettekl, klleyvraz, kl, yang,
yangetal, cherneetal} than that near the coexistence curve in systems with $R \sim 1$~\cite{langer,gunt}. 

In addition, the early
stage growth of the peak of the equal time structure function during
con\-tin\-uous ordering and spinodal decomposition in systems with $R \gg 1$ is
described by the Cahn-Hilliard-Cook (CHC) theory~\cite{cahn,hilliard,cook}
for a time proportional to $\ln R$ after the quench~\cite{binder}.
The morphology of the early stage evolution differs
from that in systems with $R\sim 1$~\cite{grossetal1, grossetal2,
klein1} for which there is no time interval when the CHC theory is applicable~\cite{gunt}.

The \mf\ limit of several earthquake fault models can be
described by an equilibrium theory~\cite{rundleetal1, kleinetal2,
fergetal}. In \nmf\ systems the smaller earthquake events are related to
fluctuations about the free energy minimum near the \psp~\cite{kleinetal2}.

In these and other examples 
the structure of the fluctuations near the
\mf\ critical point and the \psp\ is important for understanding the 
behavior of the system. In this paper we analyze 
the structure of the fluctuations and its relation to the underlying
clusters. We use field theory, scaling arguments, and cluster analysis and relate the structure of the fluctuations to the nature
of nucleation, the possible existence of a \psp\ in supercooled fluids, and
the behavior of the models of earthquake faults. The results of simulations done to test the
predictions are also discussed.

In Secs.~\ref{sec2} and
\ref{sec3} we discuss the \lg\
theory~\cite{LandauG} and the \ps~\cite{ps1,ps2} approach based on the
Langevin equation with random Gaussian noise to study fluctuations near \mf\
critical points. We use the same field theory
techniques in Sec.~\ref{sec4} to discuss fluctuations near the \sp. In
Sec.~\ref{sec6} we discuss the fluctuation morphology for \mf\ and
\nmf\ systems. In Sec.~\ref{sec7} we use the Landau-Ginzburg and \ps\
approaches to discuss the relation of the fluctuations to the clusters.
We examine the relation between the fluctuation structure
and spinodals in supercooled fluids in
Sec.~\ref{sec8} and discuss the
relation between the fluctuation structure and nucleation in Sec.~\ref{sec9}. In
Sec.~\ref{sec10} we relate the cluster structure to cellular automata models of earthquakes. We
summarize our results and discuss future work in Sec.~\ref{sec11}. The mapping of thermal systems onto 
percolation models is discussed in Appendix~\ref{sec5}.

Our main results include the following. (1) There exist objects, which we call \fc s, that have a density and lifetime dependence that is very different from the scaling of the density and lifetime of the fluctuations. The \fc s are defined by the mapping of the critical point (or spinodal) onto a percolation transition.
This difference is in contrast to non-\mf\ systems where the clusters are geometrical realizations of the fluctuations. (2) The \fc s are physical objects that have measurable consequences, which are explored for earthquake fault models, nucleation, and the measurement of the (pseudo)spin\-odal in \nmf\ systems.

\section{\label{sec2}Scaling of order parameter
fluctuations} We first discuss \nnmf\ systems
from the perspective of field theory based 
on the \lgw\ Hamiltonian~\cite{Amit, ma} 
\begin{equation}
\label{lgwh1}
H(\phi) = \!\int\!d\xv\Big[{R^{2}\over 2} [\nabla \phi(\xv)]^{2} +
\epsilon
\phi^{2}(\xv) + \phi^{4}(\xv) - h\phi(\xv)\Big].
\end{equation}
Without loss of generality we have set the proportionality
constant in Eq.~\eqref{range} equal to one. The partition function $Z$ is 
\begin{align}
\label{part}
Z &= \!\int\!\delta \phi\,e^{-\beta H(\phi(\xv))}, \\
\noalign{\noindent and the probability of the order parameter density $\phi(\xv)$ is}
\label{prob}
\pb(\phi) &= {e^{-\beta H(\phi)}\over Z}.
\end{align}
where $\beta = (k_{B}T)^{-1}$, $T$ is the absolute temperature, and
$k_B$ is Boltzmann's constant.

Because we are interested in \nnmf\ systems, we scale all lengths with $R$. We first discuss the critical point and defer our discussion of the spinodal to Sec.~\ref{sec4}. To study the critical point we set
$h = 0$ and assume that 
\begin{equation}
\label{epsilon}
\epsilon = {T - T_{c}\over T_{c}} \ll 1,
\end{equation} 
where $T_{c}$ is the critical temperature. For $\epsilon \ll 1$ and $h = 0$
we can use scaling arguments. It is straightforward to see from
Eq.~\eqref{lgwh1} that 
\begin{equation}
\label{lgwscale}
H(\t \phi) = R^{d}|\epsilon|^{2 - d/2}\!\int\!d\yv\Big[{1\over 2}\big[{\t
\nabla}{\t
\phi}(\yv)\big]^{2} \pm {\t \phi}^{2}(\yv) + {\t \phi}^{4}(\yv)\Big],
\end{equation} 
where $\yv = \xv/R\epsilon^{-1/2}$, ${\t
\phi}(\xv) = \epsilon^{-1/2}\phi(\xv)$, ${\t \nabla} = R\nabla$, and the $+$ ($-$) sign corresponds to $\e > 0$ ($\e < 0$). We take $\epsilon>0$
in this section except where otherwise noted.

The integral in Eq.~\eqref{part} can be evaluated using saddle
point techniques for $\gp \gg 1$. We can give this requirement a physical meaning from
the Ginzburg criterion~\cite{LandauG}, which states that a system can be considered to be \mf\ if
the mean square fluctuations of the order parameter are small compared
to the square of the order parameter~\cite{ma}. The order parameter $\phi$ is given by $L^{-d}\!\int\! d\xv\,\phi(\xv)$, where $L$ is the linear dimension of the system and $\phi$ corresponds to the magnetization in the Ising model.

The correlation length $\xi$ is proportional to
the linear spatial extent of the order parameter fluctuations. The
mean square fluctuations in the order parameter are characterized by
$\xi^{d}\chit$, where $\chit$ is the isothermal
susceptibility~\cite{ma}. Near a \mf\ critical point we have~\cite{stanley}
\begin{subequations}
\label{mfex}
\begin{align}
\xi & \sim R \epsilon^{-1/2} \label{mfex.a}\\
\phi &\sim |\epsilon|^{1/2} \qquad (\epsilon<0) \label{mfex.b}\\
\chit & \sim \epsilon^{-1}. \label{mfex.c}
\end{align}
\end{subequations}
(Equation~\eqref{mfex.b} is derived following Eq.~\eqref{cpopscal}.) The
Ginzburg criterion requires that 
\begin{equation}
\label{gins1}
{\xi^{d}\chit\over \xi^{2d}\phi^{2}} \rightarrow 0.
\end{equation}
If we substitute the scaling forms in Eq.~\eqref{mfex} into
Eq.~\eqref{gins1}, we obtain~\cite{binder}
\begin{equation}
\label{gins2}
G = R^{d}\epsilon^{2 - d/2} \to \infty.
\end{equation}
We will refer to 
$G = R^{d}\epsilon^{2 - d/2}$ as the Ginzburg parameter. In the limit
$G\rightarrow \infty$ the system is \mf. The system is \nmf\ for $G
\gg 1$ (but finite). The latter criterion implies the well known result that the upper
critical dimension at the critical point above which the
system has 
\mf\ critical exponents for all $R$, including $R\sim 1$, is
four~\cite{ma}. 

From Eqs.~\eqref{prob} and \eqref{lgwscale} we have
\begin{equation}
\label{probscale}
\pb(\tp) = \frac{\exp\big\{\!-\beta\gp\!\int\! d\yv [\half[{\t
\nabla}\tp(\yv)]^{2} + \tp^{2}(\yv) + \tp^{4}(\yv)\rbrack\big\}}{Z}.
\end{equation}
For $G = \gp \gg 1$ the 
Hamiltonian in Eq.~\eqref{lgwh1} can be approximated by a Gaussian~\cite{ma}:
\begin{equation}
\label{probgauss}
P_{G}(\tp) = {\exp\big\{\!-\beta\gp \!\int\!d\yv \half[{\t \nabla}
\tp(\yv)]^{2} + \tp^{2}(\yv) \big\}\over Z_G},
\end{equation}
where $Z_G$ is the functional integral over $\tp$ of the numerator in Eq.~\eqref{probgauss}. We use $P_{G}(\tp)$ to calculate the structure
function $S(\tk)$:
\begin{equation}
\label{struct}
S(\tk) = \e \lb\tp(\tvk)\tp(-\tvk) \rb = {\e \!\int\!\delta \tp(\tvk) \exp
[-\beta \gp\!\int\!d\tvk (\tk^{2} +
1)\tp(\tvk)\tp(-\tvk)\rbrack
\tp(\tvk)\tp(-\tvk)\over Z_G},
\end{equation}
where $\tvk = R\e^{-1/2}\kv$ and $\tilde \phi(\kv) = \tilde \phi(-\kv)$.
For
$\e > 0$ and $h = 0$, $\lb \t \p(\xv) \rb = 0$. We have
\begin{equation}
\label{structint}
S(\tk) \propto \frac{\e}{\gp}{1\over \tk^{2} + 1}.
\end{equation}

The Fourier transform of Eq.~\eqref{structint} gives the pair distribution
function, which we write in terms of unscaled variables:
\begin{equation}
\label{corr}
\rho^{(2)}(x) \sim \frac{1}{(x/\xi)^{d-2}} \frac{\e}
{\gp}e^{-x/\xi}.
\end{equation}
The $(x/\xi)^{2-d}$ dependence in Eq.~\eqref{corr} is valid for $d \geq 3$.
For $d=2$ this dependence is replaced by $(x/\xi)^{-1/2}$; in $d=1$
there is no $x$-dependence in the denominator. The integral
$\int\!d\xv\,\rho^{(2)}(x)$ is proportional to 
$(\epsilon/R^{d}\epsilon^{2-d/2})\xi^{d}$ for all $d$. For scaling
purposes we can treat
$\rho^{(2)}(\xv)$ for $x \lesssim \xi$ as a constant.
Because
$\rho^{(2)}(\xv \lesssim \xi)$ is proportional to the square of the density of a
fluctuation, we see that the fluctuations in the order parameter density
scale as
\begin{equation}
\label{phiscale}
\fluc(\xv \lesssim \xi) \sim {\e^{1/2}\over ({\gp})^{1/2}} = {\e^{1/2}\over
G^{1/2}}.
\end{equation}
Note that the density of 
a critical phenomena fluctuation does not scale as $\e^{1/2}$ 
as might be expected from a simple extension of 
how the order parameter scales for $\e<0$ in
Eq.~\eqref{mfex.b}.
We will discuss this point more fully in Sec.~\ref{sec4}.
The scaling of $\fluc(\xv)$ with
$G^{-1/2}$ in Eq.~\eqref{phiscale} justifies the neglect of the $\p^{4}$ term
in Eq.~\eqref{probscale} and the Gaussian approximation in
Eq.~\eqref{probgauss}. 

The susceptibility $\chi$ is related to the pair distribution function $\rho^{(2)}$ by~\cite{ma,stanley}
\begin{equation} 
\label{susc}
\chit \propto\!\int\!d\xv\,[\rho^{(2)}(\xv) - \phi^2],
\end{equation}
If we use the scaling form \eqref{phiscale} of $\fluc(\xv)$ in
Eq.~\eqref{susc}, that is, $\chit \sim \fluc(\xv)^2 \xi^d$, we find $\chit \sim \e^{-1}$, consistent with 
Eq.~\eqref{mfex.c}.

To show that $\fluc$ and $\phi$ have similar scaling behavior in a system with
$R \sim 1$ and
$d<4$, we again assume that $\rho^{(2)}(\xv \lesssim \xi)$ is a constant so that we can write Eq.~\eqref{susc} as $\chit \sim \phi_f^2 \xi^d$ and $\phi_f^2 \sim \e^{-\gamma}\e^{d \nu}$. Hyperscaling~\cite{ma,stanley} gives $\gamma + 2\beta = d\nu$ so that $\phi_f^2 \sim \e^{2\beta}$ and hence $\phi_f \sim \e^\beta$.

We next discuss the \lg\ and Cahn-Hilliard-Cook equations in
\nnmf\ systems. We can obtain these equations by noting that
the time rate of change of an order parameter such as the density is
related to the chemical potential $\mu$. If the order parameter is not
conserved and there are no other conservation laws (model A in the
Hohenberg-Halperin classification scheme~\cite{hh}), then
\begin{equation}
\label{lge1}
{\partial \phi(t)\over \partial t}\propto -\mu\ \mbox{and}\ \mu\propto
{\delta F(\phi)\over \delta \phi},
\end{equation}
where $F(\phi)$ is the free energy. We take $F(\phi)$ to be equal to the
\lgw\ Hamiltonian in Eq.~\eqref{lgwh1}, which 
is correct for \mf\ systems and a good approximation in \nmf\ systems, and assume that the
relations in Eq.~\eqref{lge1} are valid in a spatial and 
time dependent context and that the functional derivatives are with respect to
$\phi(\xv,t)$. In this way we
obtain the \lg\ equation~\cite{gunt, LandauG}:
\begin{equation}
\label{lge2}
{\partial \phi(\xv,t)\over \partial t} = -M_A\lbrack
-R^{2}\nabla^{2}\p(\xv,t) +2\e\p(\xv,t) +4{\p}^{3}(\xv,t) -
h\rbrack + \eta(\xv,t),
\end{equation}
where we have added a noise term $\eta(\xv,t)$. In the remainder 
of this section $\epsilon$ can be either positive or negative. For a
conserved order parameter (model B~\cite{hh}) we have
\begin{equation} 
\label{che1} {\pa\p(\xv,t)\over \pa t}\propto \nabla\cdot{\bf J}\
\mbox{and}\ {\bf J}\propto \nabla\mu(\xv,t).
\end{equation}
If we again interpret the right-hand side of Eq.~\eqref{lgwh1} as a free
energy and include a noise term, we obtain the Cahn-Hilliard-Cook equation~\cite{gunt}
\begin{equation} 
\label{chce1}
{\pa \p(\xv,t)\over \pa t} = M_B\nabla^{2}\lbrack -R^{2}\nabla^{2}\p(\xv,t) +
2\e\p(\xv,t) + 4{\p}^{3}(\xv,t ) - h\rbrack + \eta_{c}(\xv,t).
\end{equation}
The
quantities $M_A$ and $M_B$ in Eqs.~\eqref{lge2} and \eqref{chce1} are
mobilities and will be discussed in
Sec.~\ref{sec3}.

To obtain Eqs.~\eqref{lge1} and \eqref{che1} we assumed local
equilibrium; that is, within the coarse grained volume used
to obtain the
\lg\ free energy~\cite{gunt}, the system comes into equilibrium on a time
scale short compared to the time scales of interest.

For the remainder of this
paper we will take $\eta(\xv,t)$ and $\eta_{c}(\xv,t)$ to be
generated by a Gaussian distribution with zero mean.
That is, $\lb\eta(\xv,t)\rb = \lb\eta_{c}(\xv,t)\rb = 0$, and 
\begin{subequations}
\begin{align}
\lb\eta(\xv,t)\eta(\xv^{\prime},t^{\prime})\rb & = k_{B}T\delta(\xv -
\xv^{\prime})\delta(t - t^{\prime}) \label{noise2}\\
\lb\eta_{c}(\xv,t)\eta_{c}(\xv^{\prime}, t^{\prime})\rb &=
k_{B}T\nabla^{2}\delta(\xv - \xv^{\prime})\delta(t - t^{\prime}).
\label{noise3}
\end{align}
\end{subequations}

We can use Eqs.~\eqref{lge2} and \eqref{chce1} to determine the time
dependence of the decay of fluctuations in \nnmf\ systems. The scaling of 
$\fluc(\xv)$ in Eq.~\eqref{phiscale} implies that the cubic term in 
Eqs.~\eqref{lge2} and \eqref{che1} can be neglected. A straightforward
calculation shows that the fluctuations decay exponentially with characteristic times that diverge as $\e^{-1}$ in model
A~\cite{ma} and $R^{2}\e^{-2}$ in model B~\cite{ma}. For the remainder of this paper we will
consider only model A.

\section{\label{sec3}Parisi-Sourlas and lifetime of fluctuations}

The Parisi-Sourlas approach~\cite{ps1,ps2} begins with the \lg\ equation.
Because the noise $\eta(\xv,t)$ in Eq.~\eqref{lge2} is Gaussian, the measure of the noise is~\cite{ps1,ps2}
\begin{equation}
\label{noisem}
P(\eta) = {\exp [ -\beta\!\int\!d\xv dt\,\eta^{2}(\xv, t)]\over
\int\!\delta\eta\exp[-\beta\!\int\!d\xv dt\,\eta^{2}(\xv,t)]}.
\end{equation}
We use Eq.~\eqref{lge2} to replace
$\eta(\xv,t)$, let $h=0$ for simplicity, and express
Eq.~\eqref{noisem} as
\begin{equation}
\label{psp1}
P(\p) \propto J(\p, \eta)\exp\!\Big\{\!\!-\beta\!\!\int\!\!d \xv
dt \Big\lbrack {\pa
\p(\xv, t)\over \pa t} + M_{A}\big(\!-R^{2}\nabla^{2}\p(\xv, t) +
2\e\p(\xv, t) + 4{\p}^{3}(\xv, t){\big)}\Big\rbrack^{2}\Big\},
\end{equation}
where the Jacobian $J(\p, \eta)$ of the transformation from
$\eta$ to $\p$ is the determinant of the operator ${\delta
\eta(\xv,t)/\delta \p(\xv,t)}$. Following Parisi and Sourlas~\cite{ps1,ps2}
we introduce the Grassman variables $\pxt$ and $\pbxt$ which satisfy the
algebra
\begin{align}
\label{ga1}
\grass^{2}(\xv,t) = \grassbar^{2}(\xv,t) = \!\int\!\!d\pxt = \!\int\!\!d\pbxt =
0, \\
\label{ga2}
\big\lbrace \pbxt \pxt + \pxt \pbxt\big\rbrace = 0,\\
\label{ga3}
\int\!\!\pxt d\pxt = \!\int\!\!\pbxt d\pbxt = 1.
\end{align}
Because the variables $\grass$ and $\grassbar$ anticommute, they are referred to as
as fermions;
$\p(\xv,t)$ is a boson. With this algebra 
we can evaluate the Jacobian in Eq.~\eqref{psp1} and write
$P(\p)$ as
\begin{equation}
\label{psp2}
P(\phi, \grass, \bpsi) = {\exp\Big\{\!-\beta\big[\!\int\!\!
d\xv dt\,(S_{B}(\p, \grass, \bpsi) + S_{F}(\p,
\grass,
\bpsi)
\big]\Big\}\over \bar{Z}},
\end{equation} 
where $\bar{Z}$ is a normalization
factor. The quantities $S_{B}$ and $S_{F}$ are given by
\begin{align}
\label{Bact}
S_{B}(\p, \grass, \bpsi) &= \!\int\! d\xv dt \Big[ {\pa \p (\xv,t)\over \pa t} +
M_{A}\big(-R^{2}\nabla^{2}\p(\xv, t) + 2\e\phxt +
4\p^{3}(\xv,t)\big)\Big]^{2}
\\
\label{Fact} S_{F}(\p, \grass, \bpsi) &= \!\int\!d\xv dt\,\pbxt \Big[
{\pa \over
\pa t} + M_{A}\big(-R^{2}\nabla^{2} + 2\e + 12\p^{2}(\xv,t)\big)\Big]\pxt.
\end{align}

We first consider $S_{B}(\p, \grass, \bpsi)$ in Eq.~\eqref{Bact}. Among
the terms found by evaluating the square of the term in brackets is the contribution
\begin{subequations}
\begin{align}
\label{supsym1}
C(\phi) &= 2M_{A}\!\int\! d\xv dt {\pa \p(\xv,t) \over \pa t}\big\lbrack
-R^{2}\nabla^{2}\p(\xv,t) + 2\e\p(\xv,t) + 4{\p}^{3}(\xv,t)\big\rbrack, \\ 
&=2M_{A}\!\int\! d\xv \!\int_{t_{I}}^{t_{F}}\! dt {\pa \over \pa t}H(\p(\xv,t)), \label{supsym2}
\end{align}
\end{subequations}
where $H$ is given by Eq.~\eqref{lgwh1} with $h = 0$ and $\phi$ is replaced by $\phi(\xv,t)$. The
integral with respect to $t$ gives
\begin{equation}
\label{supsym3}
C(\p) = 2M_{A}\!\int\!d\xv \big[ H(\p(\xv,t_{F})\big) -
H\big(\p(\xv,t_{I}))\big].
\end{equation}
Parisi and Sourlas assume that $t_{F}$ and $t_{I}$ can be found such
that $C(\p) = 0$ and show that with this assumption
there is a transformation that maps fermions
and bosons into each other and keeps $P(\p)$ in Eq.~\eqref{psp2}
invariant. They refer to such systems as supersymmetric. If a system is in
equilibrium, such values of $t_{I}$ and
$t_{F}$ can always be found~\cite{ps1,ps2}.

The 
supersymmetric form of $P(\p, \grass, \grass)$ is the proper
representation for investigating
the morphology of the fluctuations in the neighborhood of \nnmf\ 
critical points. Near the latter the
Hamiltonian in Eq.~\eqref{lgwh1} can be assumed
to be Gaussian. This assumption 
implies that the \lg\ and \chc\ equations can be linearized for $G \gg 1$.
Because the $\p$ dependence in the fermionic contribution to the action
$S_{F}(\p,
\grass, \bpsi)$ comes from the nonlinear term in Eq.~\eqref{lge2}, linearization makes the two
contributions to the action, 
$S_{B}(\p, \grass, \bpsi)$ and $S_{F}(\p, \grass, \bpsi)$, independent. The
integration over the fermionic variables 
$\pxt$ and $\pbxt$ can be done immediately resulting in the measure
\begin{equation}
\label{supsym4}
P(\p) = {\exp\big\{\!-\beta \!\int\!d \xv dt\,\big({\pa \p(\xv,t)\over
\pa t}\big)^{2} + M_{A}^{2}[ -R^{2}\nabla^{2}\p(\xv,t) 
+\e\p(\xv,t)]^{2} \big\} \over Z_{S}},
\end{equation}
where $Z_{S}$ is the functional integral
over $\p$ of the numerator in Eq.~\eqref{supsym4}.

In equilibrium $P(\fluc)$ should give the same probability of a
fluctuation as $P_B(\fluc)$ in Eq.~\eqref{prob}. 
To understand the relation between these two probabilities 
we note that if $P_B(\fluc)$ is of order $e^{-1}$, then $P(\fluc)$ should
also be of order $e^{-1}$. (This requirement follows from the fact that we
expect the probability of variations from equilibrium to decay exponentially.) If we take $\fluc$ in
Eq.~\eqref{supsym4} to describe an equilibrium fluctuation,
we expect that
\begin{equation} 
\label{scal1}
\int\! d\xv dt \Big({\pa \fluc(\xv,t)\over \pa t} \Big)^{2} \sim A.
\end{equation}
where $A$ is a constant independent of $\fluc$. Without loss of generality
we can set $A = 1$.
Because the spatial
extent of $\fluc$ scales as the correlation length $\xi$, we have
from simple scaling arguments that 
$d\xv$ in Eq.~\eqref{scal1} scales as $\xi^{d}$.
If we use the scaling of $\fluc(\xv)$ in Eq.~\eqref{phiscale} and the scaling of $\xi$ in Eq.~\eqref{mfex.a}, Eq.~\eqref{scal1} implies
\begin{align} 
\label{scal2}
\frac{\fluc^2\,\xi^d}{\tfc} & \sim {\e\,R^{d}\e^{-d/2}\over
\gp\,\tfc}
\sim 1, \\
\noalign{\noindent or}
\label{csdcp}
\tfc & \sim \e^{-1}. \qquad \mbox{(lifetime of fluctuations)}
\end{align} 
Equation~\eqref{csdcp} is the well known scaling
relation for critical slowing down near
\mf\ critical points for model A~\cite{hh}.

If we
require that 
$\int\!d\xv dt\, M_A^{2} \big[ -R^{2}\nabla^{2}\fluc(\xv,t) +
\e\fluc(\xv,t)\big]^{2} \sim 1$ (see Eq.~\eqref{supsym4}),
and use the scaling relations for $\xi$, $\fluc$, and $\tfc$ and the same
arguments used to obtain Eq.~\eqref{scal2}, we find
\begin{equation}
\label{scal4}
{M_A^{2}\e^{3}R^{d}\e^{-d/2}\e^{-1}\over \gp} \sim 1,
\end{equation}
which implies that $M_A$ is a constant of order 1. 

These results are all expected. Note that
there is a significant conceptual difference between 
$P_B(\p)$ and $P(\p)$. The quantity $P_B(\p)$ is the fraction of
independent members of an ensemble in which $\p(\xv)$ is realized when a
measurement is made. Because the system is in equilibrium, we
can divide a time sequence of measurements into independent segments that
can be thought of as members of an ensemble. These segments have a duration
of the order of the decorrelation time (or longer), which is of
order $\tfc$ near the critical point. The quantity $P(\p)$
gives the probability of a ``path'' in $\p$ space. The paths of interest here are those whose probability is the order of
$e^{-1}$. The path that results in an
object with a density difference from the background of magnitude $\fluc
=\e^{1/2}/G^{1/2}$, spatial extent of order $\xi=R\e^{-1/2}$, and a
lifetime of order $\e^{-1}$ is one such path.

Suppose that in equilibrium there is an object with spatial extent 
$\xi = R\e^{-1/2}$ but a different density. In particular, suppose there is
an object of density of $\pfc \sim \e^{1/2}/G$. For reasons that
we will discuss in Sec.~\ref{sec6} we will call this object a \fc. 
Because we have assumed equilibrium, we have supersymmetry, and the action
is the sum of two contributions as in Eq.~\eqref{supsym4}. One term has the
form as in Eq.~\eqref{scal1}:
\begin{equation}
\label{act1}
S_{B,\,1} = \!\int\! d\xv dt \Big({\pa \p(\xv,t) \over \pa t}\Big)^{2}.
\end{equation}
We use reasoning similar to that following Eq.~\eqref{scal1} and define
the lifetime of the
\fc\ to be given by
$S_{B,\,1}\sim 1$. If we
substitute $\pfc \sim \e^{1/2}/G$ in Eq.~\eqref{act1}, we obtain
\begin{align}
\label{scal5}
\frac{\pfc^2\,\xi^d}{\tobjc} & \sim {\e R^{d}\e^{-d/2} \over
(\gp)^{2}\,\tobjc}
\sim 1,\\
\noalign{\noindent or}
\label{tfccp}
\tobjc & \sim {\e^{-1}\over \gp} = \frac{\e^{-1}}{G}. \qquad \mbox{(lifetime
of \fc)}
\end{align}
Equation~\eqref{tfccp} gives the lifetime of an object (the \fc) with 
density difference from the background $\pfc \sim \e^{1/2}/G$ near a
mean-field critical point.

The Boltzmann probability $P_B(\p)$ of finding such an object is
\begin{equation}
\label{boltz1}
P_B(\pfc)\propto \exp\!\Big\{\!-\beta\!\!\int\!d\xv \Big[
\frac{R^2}{2}[\nabla \pfc(\xv)]^{2} + \e
{\pfc}^{2}(\xv)
\Big]
\Big\}.
\end{equation}
If we use the scaling relations and set $\beta = 1$ for convenience, we
obtain
\begin{equation}
\label{boltz2}
P_B(\pfc)\propto e^{-1/\gp}.
\end{equation}
We see that $P_B(\pfc) \neq P(\pfc)$ despite the fact
that the system is in equilibrium. If we want the probability $
P(\pfc)$ that there is a path that consists of an object (the \fc) with
density
$\pfc\sim
\e^{1/2}/G$, spatial extent $\xi\sim R\e^{-1/2}$, and lifetime
$\e^{-1}$ rather than
$\epsilon^{-1}/G$, then using Eq.~\eqref{supsym4} we have
\begin{equation}
\label{boltz3}
P(\pfc) \sim e^{-1/\gp},
\end{equation}
and $P_B(\pfc) = P(\pfc)$.
The implication of these results is that the Boltzmann probability
$P_B(\pfc)$ requires a given time, that is, the decorrelation time, which
near a
\mf\ critical point scales as $\e^{-1}$. The probability $P(\pfc)=
P_B(\pfc)$ only if $t$ is chosen to be the decorrelation time
$\tfc$. In general, the decorrelation time is not equal to the
lifetime of the object of interest. Note that the same arguments 
apply to the normalization factors $Z$ in Eq.~\eqref{probscale} and $Z_{S}$ 
in Eq.~\eqref{supsym4}. In particular, $Z = Z_{S}$ only if the time scale
is chosen to be $\tau\sim \epsilon^{-1}$.

The mobility $M_A$ need not be a constant
independent of $\e$~\cite{gunt}. For an object with density 
$\phi_{\rm fc}\sim \e^{1/2}/G$, the second term in the action in
Eq.~\eqref{psp1} has the form
\begin{equation}
\label{act2}
S_{2} = -\!\int\!d\xv dt\, {M_{A}}^{2}\big[ R^2 [\nabla
\pfc(\xv,t)]^{2} + 2\e\pfc(\xv,t)\big]^{2}.
\end{equation}
If we use the scaling relations and the lifetime given by
Eq.~\eqref{tfccp}, we obtain
\begin{align}
\label{mob1}
\frac{ {M_A}^2 \e^{3}R^{d}\e^{-d/2}\e^{-1} } { (\gp)^{3}} & \sim 1,\\
\noalign{\noindent or}
\label{mob2}
M_{A}\sim \gp & \gg 1.
\end{align}
$M_{A}$ in Eq.~\eqref{mob2} depends on $\e$ in contrast to the
mobility in Eq.~\eqref{scal4}. If we had considered a lifetime of
$\e^{-1}$ rather than $\e^{-1}/G$ in Eq.~\eqref{tfccp}, $M_A$
would be order unity.

In summary, the probabilities $P_B(\p)$ and $P(\p)$ are equal if
the lifetime of an object is the order of $\e^{-1}$ near the \mf\ critical
point. The lifetime of an object is obtained by requiring that
$P(\p)\sim e^{-1}$. For objects with density $\e^{1/2}/G$ and a lifetime of $\e^{-1}/G$, 
the usual Boltzmann factor will not give the probability of observing such
an object.

\section{\label{sec4}Spinodals and Pseudospinodals}

In this section we discuss the meaning of spinodals and \psp s. We
begin with the Hamiltonian in Eq.~\eqref{lgwh1} and the partition
function in Eq.~\eqref{part}. For 
$G \gg 1$ the partition
function can be
evaluated using saddle point techniques, and the free energy has the Landau-Ginzburg form~\cite{LandauG},
\begin{equation}
\label{lgfe}
F = \!\int\! d\xv \,\Big[ {R^{2}\over 2}[\nabla \phi(\xv)]^{2} + \e
\phi^{2}(\xv) + \phi^{4}(\xv) -h\phi(\xv)\Big].
\end{equation}
We set the gradient term equal to zero to obtain the
free energy density
\begin{equation} 
\label{lgfe2}
f = \e \phi^{2} + \phi^{4} - h\phi.
\end{equation}
For $\e > 0$ there is only one real extremum of the free energy. For
$\e < 0$ 
there are three real extrema, one maximum and two minima. For $h =
0$ there are two states or values of $\phi$ with the
same free energy. As $|h|$ is increased, one of the minima
becomes higher than the other. The higher minimum corresponds to the metastable state. 
Increasing $|h|$ further eventually results in the
disappearance of the metastable minimum. This value of $|h|$ is referred to
as the spinodal field
$h_{s}$. It is easy to see from Eq.~\eqref{lgfe2} that at $h=h_{s}$, $f$ has
an inflection point at $\phi=\phi_{s}$. If we set (for $\epsilon<0$)
$\partial f/\partial \phi = \partial^{2}f/\partial \phi^{2} = 0$, we find
\begin{equation}
\label{spfphi}
\phi_{s} = (|\epsilon|/6)^{1/2}\ \mbox{and}\ h_{s} = 4|\epsilon|^{3/2}/(3 \sqrt 6).
\end{equation}
We define the new variable $\Delta h$,
\begin{equation}
\label{psip}
\dhs = h_{s} - h,
\end{equation}
and the new field, $\orders(\xv) = \phi(\xv) - \phi_{s} + a$, and write the
\mf\ free energy as 
\begin{equation}
\label{lgwfepsi}
F =\!\int\!d\xv \Big[ {R^{2}\over 2}[\nabla
\orders(\xv)]^{2} + \dhs^{1/2}\lambda_{1} \orders^{2}(\xv) -
\lambda_{2} \orders^{3}(\xv) + \lambda_{3} \orders^{4}(\xv)\Big].
\end{equation}
The parameter $a$ is chosen so that the
term linear in $\orders(\xv)$ does not appear in $F$. The coefficients
$\lambda_{i}$ are functions of $\epsilon$ and independent of $\dhs$. (More precisely, the $\lambda_{i}$ are a function of $\dhs$, but as $\dhs \rightarrow
0$, the 
$\lambda_{i}$ approach constants.) The free energy in Eq.~\eqref{lgwfepsi} is constructed so that
the spinodal is at 
$\orders = 0$ and $\dhs = 0$.

As for the critical point we assume that the fluctuations associated
with the spinodal can be described by a 
Gaussian-Landau-Ginzburg-Wilson Hamiltonian with the partition function
\begin{equation}
\label{lgwspin}
Z = \!\int\!\!\delta \orders \exp\big[\!-\beta \!\int\!d\xv{R^{2}\over
2}[\nabla \orders(\xv)]^{2} + \dhs^{1/2}\lambda_{1} \orders^{2}(\xv)
\big].
\end{equation}
If we follow the same argument that we used at the critical point in
Sec.~\ref{sec2}, we obtain
\begin{subequations}
\label{56}
\begin{align}
\xi & \sim R \dhs^{-1/4} \\
\chit & \sim
\dhs^{-1/2}. \\
\noalign{\noindent The fluctuations of the order parameter density scale
as}
\label{spden}
\orders_f(\xv \lesssim \xi) & \sim {\dhs^{1/2}\over \lbrack R^{d}\Delta
h^{3/2 - d/4}\rbrack^{1/2}} = \frac{\dhs^{1/2}}{G_s^{1/2}},
\end{align}
\end{subequations}
where the Ginzburg parameter near the spinodal is 
\begin{equation}
\label{ginssp}
G_{s} = R^{d} \dhs^{{3/2} - d/4}.
\end{equation}
The system is mean-field when
$G_{s}\rightarrow\infty$ and \nmf\ for $G_{s} \gg 1$. A \sp\ is only
a true critical point in the \mf\ limit~\cite{Heermann, Novotny, Gulbahce}. The distinction between a \sp\ ($G_s \to \infty$) and a \psp\ ($G_s \gg 1$) will be clear from the context.

As discussed in Sec.~\ref{sec2} the scaling of the order parameter density
and the density of the order parameter fluctuations is not
the same in
\nnmf\ systems. For example, near the critical point for $\e < 0$ we
have from Eq.~\eqref{lgfe2} 
\begin{equation}
\label{cpopscal}
-2|\e|\phi + 4\phi^{3} = 0.
\end{equation}
Equation~\eqref{cpopscal} gives the order parameter density at the minima
for $\e < 0$. As
$|\e|\rightarrow 0$ we have 
\begin{equation}
\label{phiscalecp}
\phi \sim |\e|^{1/2}.
\end{equation}

Near the spinodal we use Eq.~\eqref{lgwfepsi} and assume that
$\orders(\xv)$ is independent of $\xv$. We have
\begin{equation} 
\label{fespsig}
f(\orders) = \dhs^{1/2}\lambda_{1}\bsig^{2} - \lambda_{2}\bsig^{3}
+ \lambda_{3}\bsig^{4}.
\end{equation}
As $\dhs\rightarrow 0$, $\bsig$ scales as
\begin{equation}
\label{bsigscal}
\bsig \sim \dhs^{1/2} \qquad \mbox{(order parameter density
near the spinodal)},
\end{equation}
where we have dropped the $\orders^4$ term in Eq.~\eqref{fespsig}.
Equation~\eqref{bsigscal} gives the scaling of the order
parameter density. As near the critical
point, the scaling of the order parameter density and the density of the fluctuations is
not the same in \mf\ systems near the \sp.

In the above discussion we kept the temperature fixed and approached the
spinodal by varying the magnetic field $h$. Alternatively, we can keep the
magnetic field fixed and approach the spinodal by varying the temperature. 
To obtain the critical exponents in the temperature variable we return to
Eq.~\eqref{lgfe2} and write $\e$ as 
\begin{equation}
\label{sptemp}
\e = {T - T_{c}\over T_{c}} = {T - T_{s}\over T_{c}} + {T_{s} - T_{c} \over
T_{c}} = \es + \Delta_s, 
\end{equation}
where $T_{s}$ is the spinodal temperature for a fixed field $h =
h_{s}$. We write 
\begin{equation}
\label{sptempscal}
{\pa^{2} f\over \pa \phi^{2}}\Big|_{\phi = \phi_{s}} = -2\es
-2|\Delta_s| + 12\phi_{s}^{2} + 24\phi_{s}\orders = 0,
\end{equation}
Because $\Delta_{s}$ and $\phi_{s}$ are on the spinodal
curve, we have 
\begin{align}
\label{sptempscal1}
-2|\Delta_{s}| + 12\phi^{2}_{s} &= 0, \\
\label{sptempscal2}
-2|\es| + 24\phi_{s}\bsig &= 0.
\end{align}
From Eqs.~\eqref{bsigscal} and \eqref{sptempscal2} we have 
\begin{equation} 
\label{spcurve}
\dhs^{1/2} \sim \es,
\end{equation}
which implies from Eq.~\eqref{spden} that as the \sp\ is approached,
the density of the order parameter fluctuations scales as
\begin{subequations}
\label{fluscat}
\begin{align} 
\bsig_f(\xv \lesssim \xi) & \sim {\es \over [\gpst]^{1/2}}. \\
\noalign{\noindent Similarly, the correlation length scales as}
\xi &\sim R \es^{-1/2}, \label{clspt} \\
\noalign{\noindent and the suceptibility diverges as}
\label{susspt}
\chit &\sim \es^{-1}.
\end{align}
\end{subequations}
Equations~\eqref{56} and \eqref{fluscat} give the critical exponents near the
spinodal in terms of $\dhs$ and $\es$.

We first discuss the application of the Parisi-Sourlas
method near the \sp.
If we construct a Landau-Ginzburg equation from the free energy in
Eq.~\eqref{lgwfepsi}, we have 
\begin{equation}
\label{lgsp}
{\partial \orders(\xv,t)\over \partial t} = -M_{A,s}\big[
-R^{2}\nabla^{2} \orders(\xv,t) + 2 \dhs^{1/2}\lambda_{1} \orders(\xv,t)
- 3\lambda_{2} \orders ^{2}(\xv,t) + 4\lambda_{3} \orders ^{3}(\xv,t)\big] + \eta(\xv,t).
\end{equation}
Because the noise $\eta(\xv,t)$ is Gaussian, we
obtain an expression for the probability of a path $\orders(\xv,t)$ of the
form 
\begin{equation}
\label{pssp}
P_{\rm sp}(\orders) = {\exp\!\big[\!-\beta\!\int\!d\xv dt\,\big({\partial
\orders(\xv,t)\over \partial t}\big)^{2} + M_{A}^{2}\big\lbrace\!
-R^{2}\nabla^{2} \orders(\xv,t) + 2\lambda_{1} \dhs^{1/2} \orders(\xv,t)\big\rbrace^{2} \big]\over Z_{\rm sp}},
\end{equation}
where we have used the linear form of the \lg\ equation and have 
assumed that the system is in metastable equilibrium, which implies
supersymmetry. Arguments
similar to those used at the critical point show that the
relaxation or decorrelation time scales as 
\begin{equation}
\label{csdsp}
\tau \sim \dhs^{-1/2}. \qquad \mbox{(decorrelation time near the
spinodal)}
\end{equation}
All considerations of the difference between equilibrium Boltzmann
probabilities and probabilities of paths are the same near spinodals and
critical points.

\begin{figure}[t]
\includegraphics[scale=0.4]{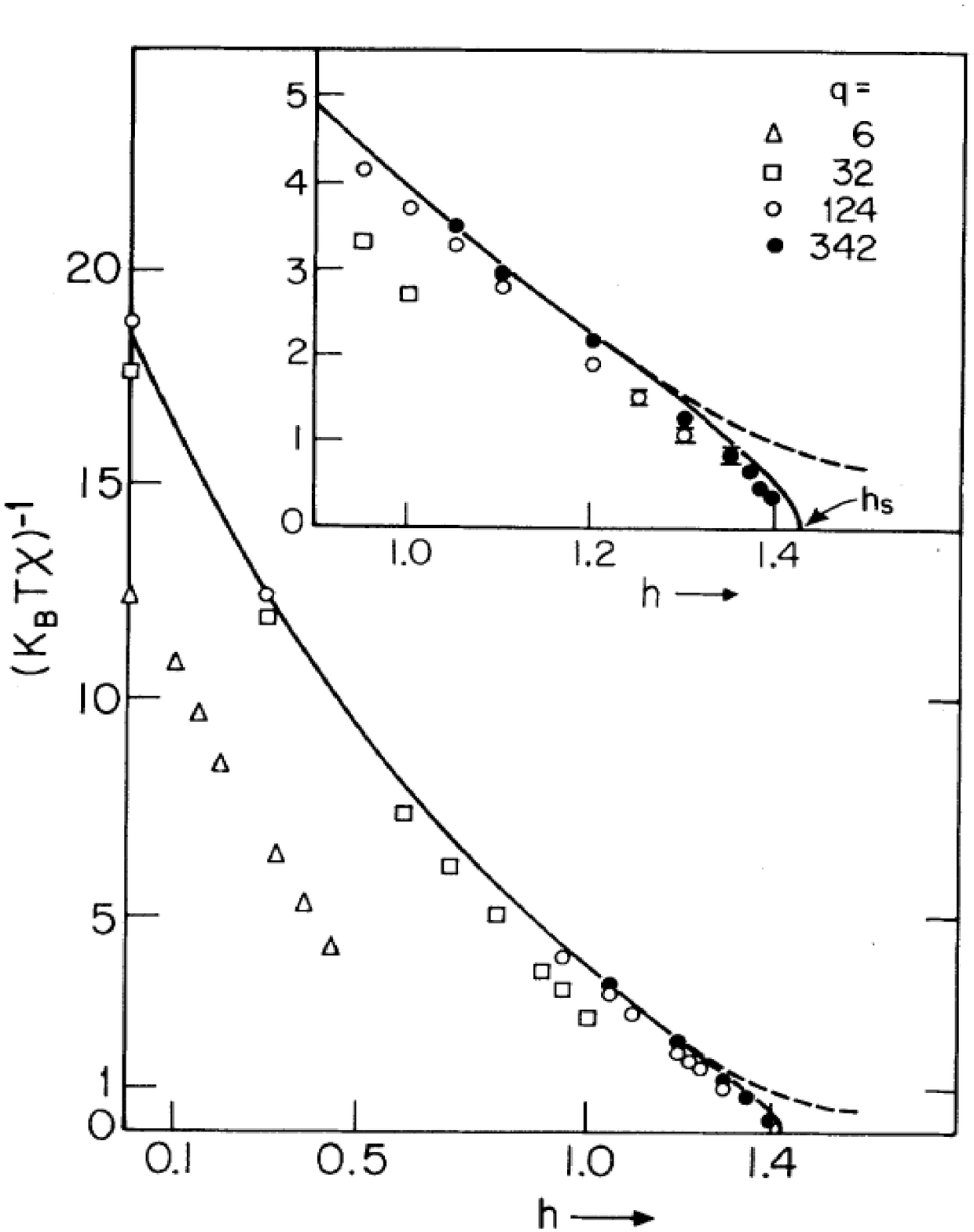}
\vspace{-0.5cm}
\caption{\label{fig1}The inverse susceptibility as a function of the
magnetic field $h$ (from Ref.~\onlinecite{Heermann}). Note that as the number of
neighbors $q$ is increased, the inverse susceptibility more closely
follows a power law. The inset shows the behavior of $(k_BT\chi)^{-1}$
closer to the \psp.}
\end{figure}

We now consider the nature of the \psp. As mentioned, for $G_s \gg 1$ but
finite, there is no spinodal. However, the system behaves as if it existed
if $G_s$ is sufficiently large. In
Fig.~\ref{fig1} we plot the inverse of the isothermal susceptibility $\chit$ found by a Monte Carlo simulation for a
$d = 3$ Ising model as a function of the applied magnetic field $h$ for
different values of $R$~\cite{Heermann}. The temperature is taken to be $4 T_{c}/9$,
where $T_{c}$ is the critical point temperature. 
The solid line is the \mf\ prediction for $q\rightarrow \infty$, where $q$ is the number of spins
that interact with a given spin~\cite{dd}. Data was taken only if the
metastable state lived longer than $10^{4}$ Monte Carlo time steps per spin.
For nearest-neighbor interactions ($q = 6$) the data stops at $h\sim 0.5$ far
from the spinodal value of the field $h_{s} = 1.43$. As 
$R$ and hence $q$ is increased, the data approaches the \mf\ result and the 
spinodal can be more closely approached. This result indicates that the larger the value of
$R$,
the more
system behaves like there is an underlying spinodal.

Another way to understand the nature of the \psp\ is to look at the
behavior of the zeros of the partition function as a function of $R$. The zeros of the partition function corresponding to the spinodal lie in
the four-dimensional complex magnetic field-temperature space for finite
$R$~\cite{Gulbahce}. As $R$ is increased, the zeros move toward the real
$(h,T)$ plane similar to the behavior of the zeros of the
partition function for Ising models in finite systems as the system size increases~\cite{ly1,ly2}. 
The idea is that the \psp\ appears to be a critical point if $h$ is not
too close to $h_s$. What is meant by too close can be
estimated by the magnitude of the Ginzburg parameter $G_{s}$ in
Eq.~\eqref{ginssp}. The value of
$\dhs$ where the spinodal concept fails can be made smaller by
increasing $R$. Hence, the theoretical arguments we made about the properties
of fluctuations near the spinodal can be tested in systems where the
interaction range $R$ is large, even though there is no true spinodal in 
nature. However, such statements have to be modified
in systems with a phase
transition that involve spatial symmetry breaking such as the liquid-solid
transition (see Sec.~\ref{sec8}).

\section{\label{sec6}Fluctuation Structure}

In this section we will use scaling arguments and the cluster mapping
discussed in Appendix~\ref{sec5} to determine the structure
of the fluctuations in 
\nnmf\ systems.

Ising critical points and the spinodal in \nnmf\ systems have been mapped onto percolation transitions~\cite{klein1, ck}. 
For simplicity, we will assume that the interaction between spins in is a constant up to a distance $R$ and is zero for distances greater than $R$.
To map the critical point onto a percolation transition we toss bonds randomly between pairs of parallel spins that are separated by a distance less than or equal to $R$ 
with a probability $p_{b} = 1 - e^{-2\beta J}$ where $J$ is the usual Ising coupling constant. This mapping guarantees that the percolation transition occurs at the critical point. For the spinodal the bond probability is $p_{s} = 1 - e^{-\beta J(1-\rho)}$, where $\rho$ is the density of spins in the stable state direction; that is, if the metastable state is in the up direction, then $\rho$ is the density of spins in the down direction. The size of the clusters is determined by the number of spins in a cluster. The details of this mapping are given in 
Appendix~\ref{sec5}. 

Equilibrium critical phenomena fluctuations in the order parameter are defined as deviations from its mean value with a linear dimension of the correlation length and a free energy cost of order 1. These properties were used in Sec.~II to calculate the density of 
fluctuations near a critical point. We will see that the same properties apply near a spinodal.

The geometrical quantity that
is isomorphic to the free energy is $-k_{B}T$ times the mean number
of clusters (see Appendix~\ref{sec5}). 
The free energy
density near the critical point
scales as $\e^{2}$, because the specific heat exponent $\alpha = 0$
for the
\mf\ critical point~\cite{ma,stanley}. Hence the free
energy $F(\e)$ in a correlation length volume scales as
\begin{equation}
\label{fe}
F(\e) \sim \e^2 \xi^d = R^{d}\e^{2 - d/2},
\end{equation}
and
the mean number of clusters in a correlation length volume scales as
\begin{equation}
\label{meannc}
\overline{n}_{\rm fc, c} \sim R^d\e^{2-d/2} = G.
\end{equation}

How are the clusters related to the fluctuations? In non-\mf\ systems 
such as the
Ising model with $R \sim 1$ and $d<4$, the mean number of
clusters in a correlation length volume near the critical point
scales as $\e^{2 - \alpha}\xi^{d} \sim 1$. The isomorphism between the Ising model and percolation implies that the the pair distribution 
function $\rho^{(2)}$ is the same as the pair connectedness function $\rho_{\rm c}^{(2)}$, which is the
probability that two spins a distance $x$ apart belong to the same cluster.
For $x \lesssim \xi$, $\rho_{\rm c}^{(2)}$ is roughly a constant and equal to $\phi^{2}_{\rm cl}$, where $\phi_{\rm cl}$
is the density of spins in the cluster. This density must be equal to the density of a
fluctuation which
scales as
$\e^{\beta}$ (see Eq.~\eqref{mfex.b}). Hence, $\phi_{\rm cl}$ scales as
$\e^{\beta}$ in a non-\mf\ system, and the clusters are a statistical realization of the
fluctuations~\cite{ck,ahrostauff}.

For $G \gg 1$ the number of clusters in a correlation length
volume near the critical point scales as
$G$ (see Eq.~\eqref{meannc}), which is much larger than unity. Are
the clusters a statistical realization of the fluctuations? To understand
that the answer is no, we note from Eq.~\eqref{phiscale} that the density of 
a critical phenomena fluctuation scales as $\fluc(\xv) \sim
\e^{1/2}/ (R^{d}\e^{2 - d/2})^{1/2}$.
If a single cluster were to correspond to a
fluctuation, then the density of spins 
in a correlation length volume would scale as 
\begin{equation}
\label{spinden}
G \fluc(\xv) = R^{d}\e^{2 - d/2}{\e^{1/2}\over
[\gp]^{1/2}} = [\gp]^{1/2}\e^{1/2}.
\end{equation}
Because the \mf\ limit corresponds to letting $R\rightarrow
\infty$ before $\e\rightarrow 0$~\cite{kacetal}, 
Eq.~\eqref{spinden} implies that the spin density is infinite in the
\mf\ limit, which is impossible. Hence, the
density of the clusters must be much smaller than the density of the fluctuations, and a fluctuation does not correspond to a single cluster as
it does in short-range systems. We conclude that the clusters in \nnmf\ 
systems play a different role, and we will refer to the clusters in these
systems as {\it fundamental clusters}.

To understand the relation between the \fc s and the fluctuations 
in a \nnmf\ system, we again use
the the fact that the pair distribution
function $\rho^{(2)}$ is isomorphic to the pair connectedness function $\rho_{c}^{(2)}$. Because the latter
is the probability that two sites a distance
$x$ apart belong to the same cluster, we have
\begin{equation}
\label{pairc}
\rho^{(2)}_{c}(\xv \lesssim \xi) \sim p_{\rm fc,c} \frac{p_{\rm fc,c}}{\gp},
\end{equation}
where $p_{\rm fc,c}$ is the probability
that the first site belongs to any one of the $\gp$ clusters, and $p_{\rm
fc,c}/\gp$ is equal to the cluster density, which is the probability that another site belongs to the same cluster as
the first.

Similarly, we have that
\begin{equation}
\rho^{(2)}(\xv \lesssim \xi) \sim \phi_f^2(\xv \lesssim \xi) \sim \Big[\frac{\e^{1/2}}{(R^d\e^{2-d/2})^{1/2}}\Big]^2.
\end{equation}
We have
$\rho_{c}^{(2)}(\xv \lesssim \xi) = \rho_{c}^{(2)}(\xv \lesssim \xi)$, and hence
We have
\begin{equation}
\label{equiv}
{p_{\rm fc,c}^{2}\over \gp} =
\bigg[{\e^{1/2}\over
\big(\gp\big)^{1/2}}\bigg]^{2},
\end{equation}
and $p_{\rm fc,c} = \e^{1/2}$. Hence, the density of spins in a \fc\ is
\begin{equation}
\label{clspns}
\phi_{\rm fc,c}(\xv \lesssim \xi) \sim \frac{p_{\rm fc,c}}{\gp} = {\e^{1/2}\over \gp} = \frac{\e^{1/2}}{G}.
\end{equation}
Because the
density of the \fc s is much smaller than the density of the fluctuations
for $G \gg 1$, a fluctuation must correspond to many \fc s.

\begin{figure}[t]
\includegraphics[scale=0.45]{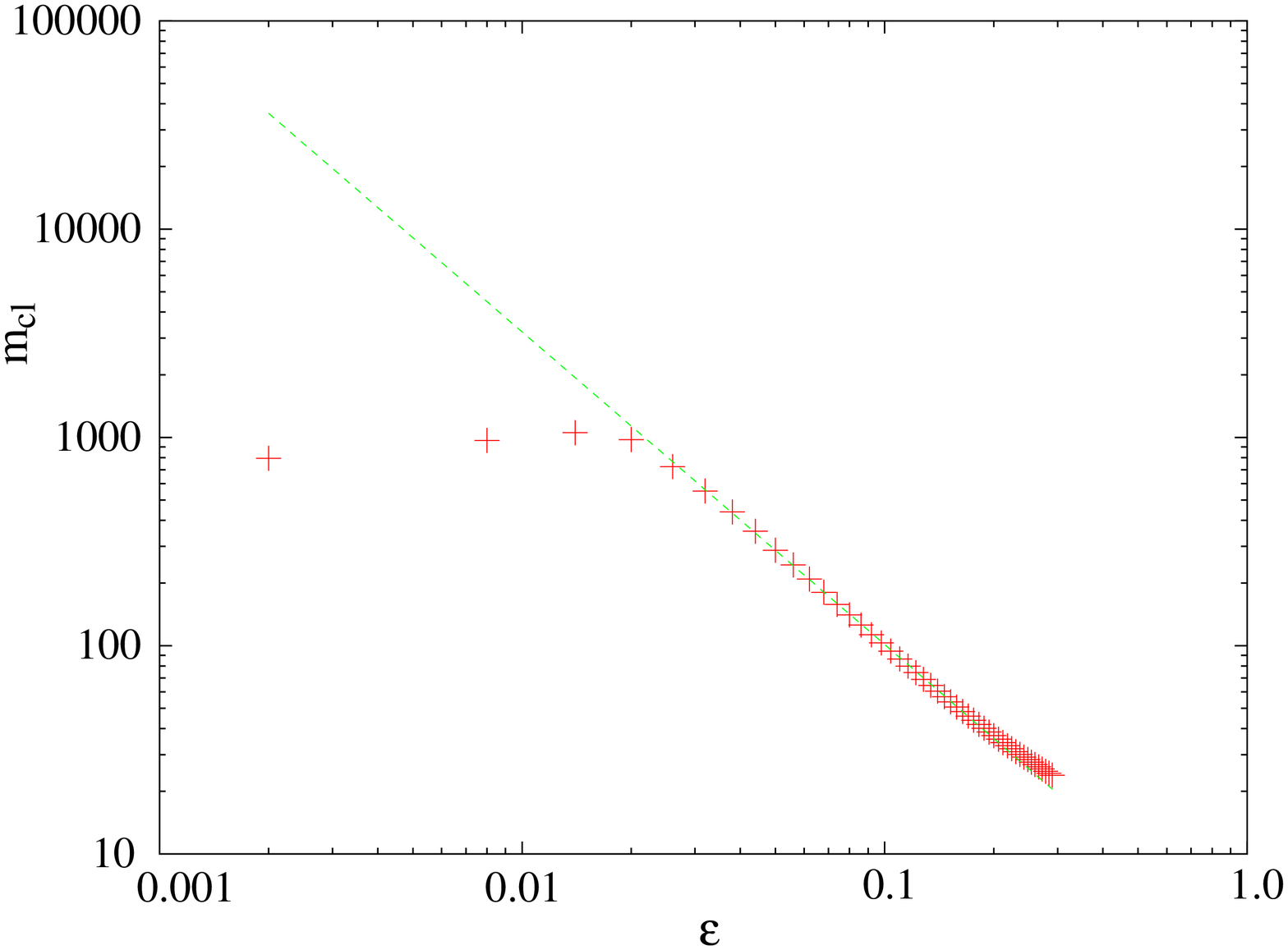}
\vspace{-0.2cm}
\caption{\label{fig2}The mean number of spins in a \fc\
with spatial extent $\xi$ found in a Monte Carlo simulation of the
$d=2$ Ising model with $R=20$ near
$T_c$ as a function of
$\epsilon$. The linear dimension of the system is $L=240$. The slope is
$\approx -1.5$ if the data is fitted in the range [0.03,\,0.2]. Note the deviation
of $m_{\rm cl}$ from the \mf\ prediction near the \mf\ critical point where
$G$ is too small to apply \mf\ arguments and where finite size effects become important.}
\end{figure}

We can test the prediction for $\phi_{\rm fc,c}(\xv)$ in Eq.~\eqref{clspns} by
determining the dependence of $m_{\rm fc,c}$, the mean number of spins in a
\fc,
on
$\e$. This dependence is given by
\begin{equation}
\label{mc}
m_{\rm fc,c} \sim \phi_{\rm fc,c}\xi^d \sim {\e^{1/2}\over \gp}R^{d}\e^{-d/2} =
\e^{-3/2}.
\end{equation}
In Fig.~\ref{fig2} we plot $m_{\rm cl}$
a function of $\e$ for fixed $R$. The slope of the log-log plot is
consistent with the theoretical prediction in Eq.~\eqref{mc}. 

We now discuss
the relation between the \fc s and the fluctuations in more detail. As discussed in Appendix~\ref{sec5} the
clusters are
constructed to be independent. Therefore a given cluster can ``flip''
independently of the other clusters. There are $\gp$ clusters near the
\mf\ critical point, half up and half down by symmetry. Because
the clusters are independent, the mean number of excess \fc s {\it in
a given direction} is determined by a random walk:
\begin{equation}
\label{deltan}
\overline{\Delta}_{\rm fc,\, c} \sim [\gp]^{1/2} = G^{1/2}.
\end{equation}
More precisely, the distribution of the number of \fc s in
a fluctuation is a Gaussian peaked about
$\overline{\Delta}_{\rm fc,\,c}$.

From this analysis we see that the density of a fluctuation is the product of the
density of
\fc s, Eq.~\eqref{clspns}, and the mean number of excess clusters,
Eq.~\eqref{deltan}:
\begin{equation}
\label{fd}
\fluc(\xv \lesssim \xi) \sim \phi_{\rm
fc,c} \overline{\Delta}_{\rm fc,\,c} 
\sim {\e^{1/2}\over \gp}(\gp)^{1/2}= {\e^{1/2}\over (\gp)^{1/2}} =
\frac{\e^{1/2}}{G^{1/2}},
\end{equation}
in agreement with Eq.~\eqref{phiscale}.

A similar analysis can be done near the spinodal, and we will only summarize the results here. In this case there is an 
infinite cluster that is a statistical realization of the metastable state
magnetization~\cite{grossetal1,grossetal2}. 
If we subtract this cluster~\cite{kleinetaltbw}, 
the results are similar to those near the \mf\ critical
point. The mean number of \fc
s in a
correlation length volume, half up and half down, is given by (compare to Eq.~\eqref{meannc})
\begin{equation}
\label{cnsp}
\overline{n}_{\rm fc,\,s} \sim \gps = G_s,
\end{equation}
and their density is (see Eq.~\eqref{clspns})
\begin{equation} 
\label{cludensp}
\orders_{\rm fc,s} \sim {\dhs^{1/2}\over \gps} = \frac{\dhs^{1/2}}{G_s}.
\end{equation}
We see that a system is
\nmf\ when the number of fundamental clusters in a correlation length volume
is large (see Eq.~\eqref{cnsp}). As
$\e$\,($\dhs$) is decreased for $d<4$ (critical point) or $d<6$
(spinodal) for fixed $R$,
$G$\,($G_{s}$) decreases and the system becomes less \mf.

If we use the same random walk argument as for Eq.~\eqref{deltan}, we find
that the density of the order parameter fluctuations near the spinodal
scales as
\begin{equation}
\label{fludensp}
\orders_{f,s}(\xv \lesssim \xi) \sim {\dhs^{1/2}\over (\gps)^{1/2}} =
\frac{\dhs^{1/2}}{G_s^{1/2}}.
\end{equation}

Hyperscaling (two exponent scaling) is satisfied for systems that
are non-\mf\ if $\e$ ($\Delta
h$) is sufficiently small~\cite{ma,stanley}. This connection suggests that
the magnitude of
$G$\,($G_{s}$) determines the existence of hyperscaling. If $G$\,($G_{s}$) decreases, the clusters must coalesce and additional length scales
are introduced~\cite{coniglio}. We now show that the existence of one relevant or divergent length scale
is insufficient for the existence of hyperscaling if $G \gg 1$ because the number of \fc s changes as the critical point is
approached. The assumption of one divergent length scale leads to the following form for the
singular part of the free energy density
near the critical point~\cite{ma,stanley}
\begin{equation}
\label{sfcp}
f(\e,h) = {1\over \xi^{d}}f(\xi^{y_{T}}\e, \xi^{y_h}h).
\end{equation}
If we differentiate $f$ twice with respect to $h$ and set $h=0$, we find
\begin{equation}
\label{sfcp2}
{\partial^{2} f(\e,h)\over \pa h^{2}}\Big|_{h=0} = {\xi^{2y_{h}}\over
\xi^{d}}{\pa^{2}\over \pa (\xi^{y_h})^{2}}f(\xi^{y_{T}}\e,
\xi^{y_{h}}h)\Big|_{h=0}.
\end{equation}
The left-hand side of Eq.~\eqref{sfcp2} is the isothermal susceptibility $\chit$. We
now fix $\xi^{y_{T}}\e$ to be equal to one. Because $f(1, h=0)$ is not
singular~\cite{ma,stanley} and
$\xi \sim \e^{-1/y_{T}}$, we have $\xi^{2y_{h} - d} = \e^{-(2y_{h} -
d)/y_{T}}$.
Hence the exponent $\gamma$ that characterizes the
divergence of $\chit$ near the critical point is given
by
\begin{align}
\label{gamma}
\gamma &= {2y_{h} - d\over y_{T}}. \\
\noalign{\noindent By using a similar argument, we obtain}
\label{beta}
\beta &= {d - y_{h}\over y_{T}}.
\end{align}
For \mf\ Ising models or simple fluids $\beta = 1/2$ and $\gamma
= 1$. 
Hence $y_{h}=3d/4$ and $y_{T} = d/2$.
For fixed $R$, $\xi =
R\e^{-1/2}$, so that $y_{T} = 2$ for all $d$~\cite{ma,stanley} and not $d/2$.
Hence, for fixed $R$ two
exponent scaling does not hold in the neighborhood of a \mf\ critical
point. The same argument holds near the spinodal.

In contrast, consider what happens for fixed $G = \gp$. From
Eqs.~\eqref{phiscale} and \eqref{susc} we have
\begin{equation}
\label{gpconst}
\chit \sim \Big\lbrack{\e^{1/2}\over
(\gp)^{1/2}}\Big\rbrack^{2}R^{d}\e^{-d/2}.
\end{equation}
If we keep $G =\gp$ constant, then
$
R^{d}\e^{-d/2} \propto \e^{-2}$. 
Therefore the susceptibility $\chit \propto \e^{-1}$ so $\gamma = 1$.
Likewise from Eq.~\eqref{fd} the
density scales as $\e^{1/2}$ so $\beta = 1/2$. For $G$ fixed $R\e^{-1/2} \propto \e^{-2/d}$ so that $\nu = 2/d$ and $y_{T} = d/2$.
Hyperscaling now holds and $\gamma + 2\beta = d\nu$. Similar arguments hold near the spinodal if the
infinite cluster is removed.

In summary, we have shown that the relation
between the clusters and the fluctuations in \nnmf\ systems is more
complex than in non-mean-field systems. In particular, the individual clusters are not realizations of the
fluctuations, which instead are
related to fluctuations of the number of clusters. For this reason we refer to the
clusters in \nnmf\ systems as \fc s. The
mean number of \fc s in a correlation length volume is proportional to the Ginzburg parameter $G$. The
dependence of $G$ on $\e$ ($G_{s}$ on $\dhs$) 
causes the breakdown of hyperscaling.

\section{\label{sec7}Lifetime of the fundamental clusters}

In Sec.~\ref{sec6} we assumed a scaling form for the free
energy. Here we do a more detailed calculation using the \lgw\
Hamiltonian in the \mf\ limit. Our main result is the scaling dependence of the lifetime of the \fc s. We also recover the same scaling results for the free energy.

If we set $h = 0$, scale all lengths with the correlation length, and assume
that $\phi(\xv)$ scales as $\e^{1/2}$ near the critical point, we obtain
the Hamiltonian in Eq.~\eqref{lgwscale}. If we assume
that
${\t \phi}(\yv)$ is independent of $\yv$
and restrict the integral to a
region the size of a correlation length volume, we have 
\begin{equation}
\label{H}
H(\phi) = R^{d}\e^{2 - d/2}[\pm {\t \phi}^{2} + {\t
\phi}^{4}],
\end{equation}
and the partition function becomes 
\begin{equation}
\label{pf}
Z(\e) = \!\int_{-\infty}^{\infty}\! d{\t \phi}\, e^{-\beta
R^{d}\e^{2 - d/2}( \pm {\t \phi}^{2} + {\t \phi}^{4})}.
\end{equation}
For $G = \gp \gg 1$ we can do the integral in Eq.~\eqref{pf} using saddle point
techniques. For $\e < 0$ the saddle points are at $\phi = \pm 1/{\sqrt
2}$, and we obtain
\begin{equation}
\label{lgwpf2}
Z(\e) \propto e^{\beta \gp},
\end{equation}
and hence the free energy is 
\begin{equation}
\label{fe1}
-k_{B}T\ln Z(\e) = -\gp. \qquad (G \gg 1)
\end{equation}
We have neglected the logarithmic corrections generated by the steepest descent integral.
Note the minus sign on the right-hand side of Eq.~\eqref{fe1}. For $\e >
0$ the saddle points are at
$\pm i/{\sqrt 2}$ and the free energy is also proportional to $-\gp$.
Hence as argued in Sec.~\ref{sec6} and Appendix~\ref{sec5} we can use the percolation mapping to show that the number of \fc s
scales as
$\gp$ near the critical point. As usual, similar arguments can be used near
the spinodal for
$G_s\rightarrow
\infty$ and near the \psp\ for $G_s\gg 1$.

To determine the lifetime of
the fundamental clusters, we return to the
Parisi-Sourlas method. 
At the spinodal we have from Eq.~\eqref{pssp}
\begin{equation}
\label{spsc}
\int\!d\xv dt \Big({\pa \orders\over \pa t}\Big)^{2} \sim 1.
\end{equation}
As in Sec.~\ref{sec3} the lifetime of a \fc\ is found by requiring that
\begin{equation}
\label{clusplt}
\frac{\orders_{\rm fc,s}^2\,\xi^d}{\tobjs} \sim {(\dhs^{1/2})^{2}R^{d} \dhs^{d/4}\over [\gps]^2\,\tobjs} \sim 1,
\end{equation}
or 
\begin{equation}
\label{tfcs}
\tobjs \sim {\dhs^{-1/2}\over \gps}. \quad \mbox{(fundamental cluster
lifetime near spinodal)}
\end{equation}
We see that near the critical point ($G \gg 1$) and the \psp\ ($G_s \gg 1$), the lifetime of the fundamental
clusters is considerably shorter than the lifetime (decorrelation time) of a
fluctuation near the critical point ($\e^{-1}$) and the spinodal
($\dhs^{-1/2}$).

To understand the relation between the lifetime of the fundamental clusters and
the lifetime of a critical phenomena fluctuation recall that the clusters are independent. We consider the
fluctuations to be formed from the ``vacuum'' (zero magnetization near the
critical point and zero net magnetization after the infinite cluster is subtracted near the spinodal) by a random walk in the number
of fundamental clusters. 
At the critical point the density of the critical phenomena fluctuations is given by Eq.~\eqref{phiscale}.
Because the fluctuations arise from a
random walk in the number of \fc s, there must be a ``walk'' of
$(\gp)^{1/2}$\ cluster flips (steps) in the direction of the fluctuation
to obtain a density of
\begin{equation}
\label{cpd2}
\phi_f(\xv) \sim {\e^{1/2}\over \gp}(\gp)^{1/2} = {\e^{1/2}\over (\gp)^{1/2}}.
\end{equation} 
The time needed for $\gp$\ attempted cluster
flips is
\begin{equation}
\label{cpd3}
\tfc \sim {\e^{-1}\over \gp}\gp = \e^{-1},
\end{equation}
in agreement with Eq.~\eqref{csdcp}. The same considerations near
the \sp\ yield 
$\tfs \sim \dhs^{-1/2}$
for the lifetime of a fluctuation, in agreement with our earlier result for $\tfs$ in Eq.~\eqref{csdsp}.

\begin{table}[t]
\begin{tabular}{|p{4.5cm}|l|l|l|}
\hline
quantity & \mf\ critical point & spinodal & hyperscaling\\
\hline
Ginzburg parameter & $G=R^{d}\epsilon^{2 - d/2}$ & $G_s=\gps$ & $G$\,($G_s$) fixed\\
\hline
order parameter & $\phi \sim \e^{1/2}$ & $\orders \sim \dhs^{1/2}$ & $\sim \e^\beta$\\
\hline
fluctuations in order parameter density & 
$\fluc(\xv \lesssim \xi) \sim \e^{1/2}/
G^{1/2}$ & $\orders_f(\xv \lesssim \xi) \sim \e^{1/2}/
G_s^{1/2}$ & $\sim \e^\beta$\\
\hline
lifetime of fluctuation & $\tfc \sim \e^{-1}$ & $\tfs \sim \es^{-1}$ & $\sim \e^{-z/\nu}$\\
\hline
lifetime of fundamental cluster & $\tobjc \sim \e^{-1}/G$ & $\tobjs \sim \es^{-1}/G_s$ & \\
\hline
mean number of clusters 
in correlation length volume & $\overline{n}_{\rm fc, c} \sim G$ & $\overline{n}_{\rm fc, s} \sim G_s$ & $\sim 1$ \\
\hline
density of fundamental cluster & $\phi_{\rm fc,c}(\xv \lesssim \xi) \sim \e^{1/2}/G$ & $\orders_{\rm fc,s}(\xv \lesssim \xi) \sim \e^{1/2}/G_s$ & \\
\hline
mean number of excess
\fc s & $\overline{\Delta}_{\rm fc,c} \sim G^{1/2}$ & $\overline{\Delta}_{\rm fc,s} \sim G_s^{1/2}$ & \\
\hline
\end{tabular}
\caption{\label{tab1} Summary of our notation and some of the important scaling relations derived in the text. The \sp\ can be approached by reducing the magnetic field difference $\dhs$ for fixed temperature or by decreasing the temperature difference $\es$ for fixed 
magnetic field. The two approaches are related by $\dhs^{1/2} \sim \es$. The exponent $z$ characterizes critical slowing down and is of order 2 for systems described by model A~\cite{hh}.}
\end{table}

In summary, we have argued that the fluctuations near a \mf\ critical point
and a spinodal are not represented by a single \fc. The relation between these clusters and the fluctuations in \nnmf\ systems is qualitatively
different than in systems that
obey hyperscaling. In the former the fluctuations are formed by a
random walk in the number of \fc s that ``flip'' on a time scale
much shorter than the scale set by critical slowing down. A summary of our notation and our main results so far is given in Table~\ref{tab1}.

\section{\label{sec8}Cluster Structure and Instabilities in 
Supercooled Liquids} 

We begin our discussion of the consequences of the fluctuation 
structure in \nnmf\ systems
by considering the liquid-solid spinodal in supercooled
fluids. To explain the role of the structure of the fluctuations we 
first provide some background. In 1951 Kirkwood~\cite{kirk} noted that
approximate equations for the distribution functions in the liquid state
appeared to show an instability as the supercooled liquid is quenched
deeper. Kirkwood began with the first equation of the static BBGKY
hierarchy~\cite{hill}
\begin{equation}
\label{bbgky2}
-k_{B}T\nabla_{1}\rho^{(1)}(\xv_{1}) = \!\int\! d\xv_{2}\,
\nabla_{1}V(x_{12})\rho^{(2)}(\xv_{1},\xv_{2}),
\end{equation}
where $\rho^{(1)}(\xv_{1})$ and $\rho^{(2)}(\xv_{1}, \xv_{2})$ are
the one and two particle distribution functions respectively, $\nabla_{1}$
denotes differentiation with respect to the position of particle $1$, and
the interaction potential $V(x_{12})$ is assumed to be pairwise
additive and spherically symmetric. Suppose 
that the system is in the liquid phase where $\rho^{(1)}(\xv_{1})$ is a constant equal to $\rho$, and $\rho^{(2)}(\xv_{1}, \xv_{2})$ is a
function of $x_{12} = |\xv_{1} - \xv_{2}|$ and is equal to
$\rho^{2}h(x_{12}) = 
\rho^{2}(1 + g(x_{12}))$,
where $g(x_{12})$ is the pair correlation
function~\cite{hill}.
We substitute
\begin{equation}
\label{kmsa}
\rho^{(1)}(\xv_{1}) = \rho +
\omega(\xv_{1})
\end{equation}
into Eq.~\eqref{bbgky2}, treat $\omega(\xv_{1})$
as a small perturbation, and linearize Eq.~\eqref{bbgky2} to find
\begin{equation}
\label{bbgky3}
-k_{B}T{\nabla_{1}\omega(\xv_{1})\over \rho} = \!\int\!
d\xv_2\,\nabla_{1}V(x_{12})h(x_{12})\omega(\xv_{2}),
\end{equation} 
where the spherical symmetry of $V(x_{12})$ and $h(x_{12})$ results in 
\begin{equation}
\label{zero}
\int\!d\xv_{2}\, \nabla_{1}V(x_{12})h(x_{12}) = 0.
\end{equation}
We have also assumed that a possible instability in $h(x_{12})$ is 
higher order in $\omega$. If we define
\begin{equation}
\label{defq}
q(\xv_{1} - \xv_{2}) = \nabla_{1} V(x_{12})h(x_{12}),
\end{equation}
we see that there is an instability~\cite{kirk} if there is a nonzero solution to
\begin{equation}
\label{kinst}
k_{B}T\nabla_{1}\omega(\xv_{1}) + \beta\rho\!\int\!d\xv_{2}\,q(\xv_{1} -
\xv_{2})\omega(\xv_{2}) = 0.
\end{equation} 
Kirkwood analyzed Eq.~\eqref{kinst} for the hard sphere
fluid and found that there was an instability in $d = 3$~\cite{kirk}. However,
he ignored a possible instability in $h(x_{12})$
which is related by the BBGKY~\cite{hill} hierarchy to possible
instabilities in all of the distribution functions. Hence, it is not clear
that the instability Kirkwood found is real. A more careful analysis~\cite{lov} suggests that the instability vanishes
when higher order terms are considered. 

To investigate the existence of an instability and its relation to a
possible spinodal, Grewe and Klein~\cite{gk1,gk2} investigated the properties
of a simple fluid for which the interaction potential has the Kac form~\cite{kacetal} given in Eq.~\eqref{kacpot} with $V_R=0$ and
\begin{equation}
\label{kacf12}
\Phi(\gamma| \xv|) = 
\begin{cases}
1 & \text{if $\gamma|\xv|\leq 1$}, \\
0 & \text{if $\gamma|\xv|$} > 1.
\end{cases}
\end{equation}
In the \mf\ limit $\gamma\rightarrow 0$, it was shown that all distribution
functions of order higher than two are completely specified by only
the single particle and pair distribution functions and that $\rho^{(1)}(\xv_{1})$
in the limit $\gamma\rightarrow 0$ satisfies
the equation~\cite{gk1,gk2}
\begin{equation}
\label{km}
\rho^{(1)}(\xv_{1}) = z\exp\lbrack-\beta\!\int\! d\xv_{2}\,\Phi(|\xv_{12}|)\rho^{(1)}(\xv_{2})\rbrack,
\end{equation}
where $z=e^{-\beta \mu}$ and $\mu$ is the chemical potential.
Similarly, $g(|\xv_{12}|)$ satisfies
\begin{equation}
\label{oz}
g(|\xv_{12}|) = \beta\rho\Phi(|\xv_{12}|) - \beta\rho\!\int\!d\xv_{3}\,g(|\xv_{1} - \xv_{3}|)\Phi(|\xv_{2} - \xv_{3}|),
\end{equation}
where all length scales are in units of $\gamma^{-1} = R$.
Note that $g(|{\xv}_{12}|)$ is of order $\gamma^{d}$. The derivations of Eqs.~\eqref{km} and \eqref{oz} are given in Refs.~\onlinecite{gk1} and \onlinecite{gk2}.

From Eq.~\eqref{oz} the structure function $S(k)$, which is obtained
by taking the Fourier transform of $g(|{\xv}_{12}|)$, is proportional to
\begin{equation}
\label{ozsf}
S(k) \propto {1\over 1 + \beta\rho{\hat \Phi}(k)},
\end{equation}
where $\hat \Phi(k)$ is the Fourier transform of $\Phi(\gamma |\xv|)$ and $k=|\kv|$.
Note that
the structure function is order one in the $\gamma \rightarrow 0$
limit. 

We can perform a stability analysis on Eq.~\eqref{km} similar to that
done by Kirkwood on Eq.~\eqref{bbgky2}. We substitute Eq.~\eqref{kmsa} into Eq.~\eqref{km} and
linearize in
$\omega(\xv_{1})$ to obtain
\begin{equation}
\label{mfins1}
\omega(\xv_{1}) = -\beta\rho\!\!\int\!d\xv_{2}\,\Phi(|\xv_{1} - \xv_{2}|)\omega(\xv_{2}),
\end{equation}
where $\rho$ is the solution of 
$
\rho = z\exp[-\beta {\hat \Phi}(0)\rho]$,
and
$\hat{\Phi}(0) = \!\int\!d\xv\,\Phi(|\xv|) > 0$.

There is an instability only if there is a nonzero solution to
Eq.~\eqref{mfins1}. If we take the Fourier transform of
Eq.~\eqref{mfins1}, we can express the instability condition as 
\begin{equation}
\label{kminst2}
1 + \beta\rho{\hat \Phi}(|\kv|) \leq 0,
\end{equation}
or $\hat{\Phi}(|\kv|)<0$ for some value of $|\kv|$. This condition is
satisfied for the potential in Eq.~\eqref{kacf12}. Because 
$\Phi(\gamma |\xv|)$ in Eq.~\eqref{kacf12} has a Fourier transform that is
bounded from below, there is a value of
$\beta\rho$ below which there is no instability. If $k_{0}=|\kv|$ is the
location of the global minimum of $\hat{\Phi}(|\kv|)<0$, then the
system has no instability for $\beta\rho < -1/{\hat
\Phi}(k_{0})$. We see from Eq.~\eqref{ozsf} that the
structure function $S(k_0)$ first diverges for fixed $\rho$ (as $T$ is decreased) at the same value of the
temperature $T$ at which an instability first appears. The divergence of
$S(k_0)$ implies that the instability in the \mf\ system is a spinodal and is analogous to the divergence of the
susceptibility at the Ising spinodal.

In the \mf\ limit no higher order distribution functions need
to be considered and the results of Grewe and Klein~\cite{gk1,gk2} are
rigorous. The structure function $S(k_0)$ diverges as $(T - T_{s})^{-1}$ so that the
critical exponent is the same as the Ising spinodal if the temperature
rather than the magnetic field is used to approach the spinodal in the
Ising model. The only difference is that $S(k)$
diverges at $k = 0$ in Ising models and gases rather than at $k_{0}\neq 0$. The other
critical exponents are also the same as for the Ising
spinodal~\cite{gk1,gk2}.

We next discuss an important
difference between measurements of the spinodal exponents
in Ising models and in supercooled fluids. As we have discussed (see Fig.~\ref{fig1}), there is no spinodal in an Ising model for $R$ finite, but we see
spinodal-like behavior for
$R\gg 1$ if the system is not quenched too deeply into the
metastable state~\cite{Heermann, Novotny,Gulbahce}.
The larger $R$, the more the \psp\ behaves like a true
spin\-odal.

In the supercooled liquid there is no direct evidence
of a spinodal or \psp\ from either experiments or simulations. We
will see that this lack of direct evidence is due to the structure of the fluctuations in \nnmf\ systems and the crucial role of the \fc s.

For the potential in Eq.~\eqref{kacf12} the system is a fluid for high
temperatures and/or low densities~\cite{gk1,gk2,mel}. If the temperature $T$
is lowered at a fixed density, the liquid-solid instability is
encountered at the spinodal temperature $T_{s}$. If
$T$ is lowered below $T_{s}$, the uniform density fluid phase becomes unstable and a ``clump''
phase is formed~\cite{mel}. Because we are interested in the nature of the
spinodal, the behavior of the system for $T < T_s$ is not of interest here.

Unlike Ising/Potts models, there is no
precise definition of a cluster in a continuum system of particle. However, it is reasonable to assume that the scaling behavior of the fluctuations and
\fc s near the liquid-solid spinodal is the same as near the Ising spinodal. This assumption is
consistent with the fact that the Ising and spinodal
exponents for the system defined by Eq.~\eqref{kacf12} are the same. For convenience, we will use temperature scaling
near the liquid-solid transition rather than the analog of magnetic field scaling. From Eqs.~\eqref{tfcs}
and
\eqref{spcurve} the lifetime of the fundamental clusters as $\es \to 0$ scales as
\begin{equation}
\label{ltfct}
\tobjs \sim {\es^{-1}\over \gpst}. \qquad \mbox{(lifetime of \fc\ near spinodal)}
\end{equation}

Grewe and Klein~\cite{gk1,gk2} used the \mf\ formalism
developed by Kac et al.~\cite{kacetal}. In particular, the interaction
range $R =
\gamma^{-1}$ is taken to infinity before the \sp\ is
approached, that is, before the limit $\es \propto (T - T_{s})$ goes
to zero. Hence, in the \mf\ limit there exists \fc s with
probability of order one but zero lifetime (see the discussion in
Sec.~\ref{sec3}). 
In experiments and simulations, the converse is true.
For example, consider a
measurement of the structure function $S(k)$ in a simulation. The structure function is obtained by computing
\begin{equation}
\label{ssf}
S(k) =\frac1N \langle \big\lbrack \sum_{j} e^{i \kv \cdot
\xv_{j}}\big\rbrack^{2}\rangle,
\end{equation}
where ${\xv}_{j}$ is the instantaneous position of particle $j$ and
$\langle \cdots\rangle$ denotes an ensemble average. Because a simulation
can be performed only on systems with finite
$R$, the time of the measurement, which is instantaneous, is much less
than the nonzero \fc\ lifetime in Eq.~\eqref{ltfct}. Therefore the
result of the simulations need not be consistent with the \mf\
predictions~\cite{gk1,gk2}. (The
time scale of the measurement does not refer to the time over which data is
taken, but to the time during which the probe (say a neutron) is in contact
with a fluctuation.)

Before we consider the behavior of $S(k_0)$
in supercooled liquids, we discuss the
measurement of $S(k)$ near the Ising spinodal. We will find that the measured
behavior of $S(k=0)$ agrees with the \mf\
predictions~\cite{dd}. The application of the scaling argument in Sec.~\ref{sec6} to the
susceptibility
near the Ising spinodal gives
(see Eqs.~\eqref{susc} and \eqref{fludensp})
\begin{equation}
\label{spscal2}
\chit \sim \orders_f^2\,\xi^d \sim \Big\lbrack{\dhs^{1/2}\over
(R^{d} \dhs^{3/2-d/4})^{1/2}}\Big\rbrack^{2} R^{d} \dhs^{-d/4}\sim
\dhs^{-1/2}.
\end{equation}
Alternatively, we can
calculate $\chit$ directly from the \fc s. The density of a fundamental cluster is
given in Eq.~\eqref{cludensp}. Hence, the isothermal susceptibility
associated with one \fc\ is (see Eq.~\eqref{fludensp})
\begin{equation}
\label{phiscagain2}
\chi_{\rm 1fc} \sim \orders_{fc,s}^2\,\xi^d\sim \Big\lbrack{\dhs^{1/2}\over R^{d} \dhs^{3/2 -
d/4}}\Big\rbrack^{2}R^{d} \dhs^{-d/4}\sim {\dhs^{-1/2}\over
R^{d} \dhs^{3/2 - d/4}}.
\end{equation}
Because the clusters are independent, the isothermal
susceptibility of the system is the sum of the individual cluster isothermal
susceptibilities. Because there are $R^{d} \dhs^{3/2 - d/4}$ clusters in a correlation length volume, $\chit=
\dhs^{-1/2}$, consistent with Eq.~\eqref{spscal2}. Although there are
clusters of both up and down spins, each cluster has the same isothermal
susceptibility because the susceptibility is proportional to the square of
the spin density~\cite{ma,stanley}.

Another way of obtaining the same result as in
Eq.~\eqref{spscal2} is to calculate the structure function $S_{\rm 1fc}(\kv)$ for one \fc\ from
the Fourier transform of the connectedness function of
the cluster. We can write (see Eq.~\eqref{cludensp})
\begin{equation}
\label{sfclust}
S_{\rm 1fc,\,s}(\kv) \sim \Big\lbrack{\dhs^{1/2}\over R^{d} \dhs^{3/2
- d/4}}\Big\rbrack^{2}\delta(\kv).
\end{equation}
The delta function comes from integrating over the infinite size
of the spatially uniform cluster.
For a cluster with the spatial extent of the
correlation length
$\xi$, $\delta(\kv)$ in Eq.~\eqref{sfclust} would be replaced by a function
whose height is
$\xi^{d}$ and whose width is proportional to $\xi^{-1}$. 

The difference between the \fc\ lifetime $\tobjs$ in Eq.~\eqref{tfcs} and the lifetime of a
fluctuation 
$\sim
\dhs^{-1/2}$ suggest the following picture. A
fluctuation is a collection of fundamental clusters that appear and
disappear on the time scale $\tobjs$. The lifetime of a fluctuation is a
much longer than a \fc\ for $G_s
\gg 1$, which implies that the fundamental clusters
come and go with different angular orientations but with their centers in
roughly the same place for a time of order $\dhs^{-1/2}$, the lifetime
of a fluctuation. If we make a measurement on a time scale
$\gg \dhs^{-1/2}$, the number of up and down fundamental
clusters would be the same on the average, and the measured structure function
would show no correlations between the spins.

Consider a measurement on a time scale $t_{\rm meas}$ such that
$t_{\rm meas} 
\gg \tobjs$ and $t_{\rm meas} \sim 
\dhs^{-1/2}$. Because the measurement time is comparable to the lifetime of the fluctuation, the measurement will see a
spin density equal to the fluctuation density, which is generated by 
fluctuations in the number of fundamental clusters. That is, the individual
fundamental clusters cannot be distinguished, and
\begin{equation}
\label{sffluct}
S(\kv) \sim \Big[{\dhs^{1/2}\over (R^{d} \dhs^{3/2 -
d/4})^{1/2} }\Big]^{2}\delta(\kv).
\end{equation}
If we replace $\delta(\kv)$ by $\xi^{d}$, we obtain the same result as in
Eq.~\eqref{spscal2}, obtained by considering the fluctuations directly.

Now suppose that a measurement is made such that $t_{\rm meas}
\ll \tobjs$.
In this case an external probe or a simulation would see a set of
order $R^{d} \dhs^{3/2 - d/4}$ {\it frozen} \fc s. To determine the
structure function that would be measured, we need to add the cluster
structure function in Eq.~\eqref{sfclust} for the
$R^{d} \dhs^{3/2 - d/4}$ frozen clusters. To do so we
convert the sum to an integral by using one of the factors of $1/
R^{d} \dhs^{3/2 -d/4}$ in Eq.~\eqref{sfclust} to create an infinitesimal
element of solid angle
$d\Omega$. (We ignore numerical factors 
because we are interested only in the scaling properties.) In so doing we
are assuming that the 
$R^{d} \dhs^{3/2 - d/4}$ fundamental clusters overlap each other with
random orientations. Hence the
sum over clusters becomes an integral over solid angle:
\begin{equation}
\label{sfint1}
S(|\kv|) \sim \!\!\int\!d\Omega\,{\dhs\over R^{d} \dhs^{3/2 -
d/4}}\delta(\kv).
\end{equation}

For $d = 3$ we have in spherical polar coordinates
\begin{equation}
\label{sfint2}
S(k) \sim {\dhs\over R^{3} \dhs^{3/2 - 3/4}}
\!\int\!\sin \theta d\theta d\phi\,
{\delta(k)\delta(\theta)\delta(\phi)\over k^{2}\sin \theta}.
\end{equation}
Hence 
\begin{equation}
\label{sfint3}
S(k) \sim {\dhs\over R^{3} \dhs^{3/2 - 3/4}}{\delta(k)\over k^{2}}.
\end{equation}
If we now replace $\delta(k)$ by $\xi$ with $k\sim
\xi^{-1}$, we obtain
\begin{equation}
\label{sfint4}
S(k)\sim {\dhs\over R^{3} \dhs^{3/2 - 3/4}}\xi^{3} = {\dhs\over
R^{3} \dhs^{3/2 - 3/4}}R^{3} \dhs^{-3/4} = \dhs^{-1/2},
\end{equation} 
in agreement with Eq.~\eqref{spscal2}. The generalization
to arbitrary dimensions is straightforward. We conclude that
we obtain the same scaling dependence for the susceptibility (equal to $S(k=0)$), independent of the relative order of magnitude of
the measurement time and consistent with simulations of the Ising model~\cite{Heermann}.

We now discuss the measurement of $S(k)$ near the liquid-solid spinodal. As stated, we
assume that the \fc s in the supercooled liquid scale the same way as
they do in Ising models. Because the clusters are independent~\cite{ck}, the
structure function of a single \fc\ has to contain information about the
symmetry of the instability. That is, a collection of
independent clusters cannot generate a symmetry that does not already
exist in each cluster. Because we are interested in
the limit of stability of the supercooled liquid and know that the
instability occurs at 
$k_{0} \neq 0$, the clusters must reflect this symmetry.
Hence, we will assume that the clusters have a symmetry reflected by
the wave vector
$\kv_{0}$ with arbitrary orientation. In analogy with
the Ising spinodal, we expect that the structure function
$S_{\rm fc,s}(\kv)$ of the \fc s can be approximated near the spinodal by 
\begin{equation}
\label{sfscl1}
S_{\rm fc,\,s}(\kv) \sim \Big\lbrack{\es\over R^{d}\es^{3 -
d/2}}\Big\rbrack^{2}\delta(\kv - \kv_{0}).
\end{equation}
We have used temperature variables rather than the
chemical potential, the analog of the magnetic field. There are other peaks in $S_{\rm fc,\,s}(\kv)$
at $|\kv| \neq \kv_{0}$, but we will focus on
$S_{\rm fc,\,s}(\kv_0)$, the peak associated with the
divergence as the liquid-solid
spinodal is approached. 

We first consider a measurement on a time scale such that
$t_{\rm meas} \ll \tobjs$, where $\tobjs$ is give by Eq.~\eqref{ltfct}.
As before we need to sum over all orientations of the frozen
clusters whose centers are fixed. We convert this sum to an
integral by absorbing one of the factors of $1/R^{d}\es^{3-d/4}$ in
Eq.~\eqref{sfscl1} to form an infinitesimal. In this case we need
to integrate over orientations of the vector
$\kv_{0}$ keeping the magnitude $k_{0}$ constant. For
$d = 3$ we have
\begin{equation}
\label{sfscl2}
S(k)\sim {\es^{2}\over R^{3}\es^{3 - 3/2}}{1\over k_{0}^{2}} \!\iint\!
d\theta d\phi\,\delta(k - k_{0})\delta(\theta)\delta(\phi).
\end{equation}
The factor of $\sin \theta$ in the numerator associated with the solid angle
and the
$\sin \theta$ in the denominator associated with the delta function in
spherical polar coordinates cancel. The integrals in Eq.~\eqref{sfscl2} give
\begin{equation}
\label{sfscl3}
S(k) \sim {\es^{2}\over R^{3}\es^{3 - 3/2}}{1\over k_{0}^{2}}\delta(k -
k_{0}).
\end{equation}
As for the Ising case we replace the delta function by the correlation length 
$\xi\sim R\es^{-1/2}$. Hence, the structure function at $k = k_{0}$ 
scales as
\begin{equation}
\label{sfscl4}
S(k_0) \sim {\es^{2} R\es^{-1/2}\over R^{3}\es^{3 - 3/2}} = R^{-2}\es^0. \qquad
(d=3)
\end{equation}
Equation~\eqref{sfscl4} implies that there
is either no divergence or the divergence is logarithmic, which is
consistent with the fact that no direct evidence of a
\psp\ has been observed in simulations of simple fluids
in $d=3$, despite the indirect evidence that nucleation is
influenced by a \psp\ in deeply quenched Lennard-Jones
liquids~\cite{yangetal, ten, trudu} and in nickel~\cite{cherneetal}.

It is easy to
show that $S(k_0)$ for arbitrary dimensions scales as
\begin{equation}
\label{sfscl4.d}
S(k_0) \sim {\es^{2} R\es^{-1/2}\over R^{d}\es^{3 - d/2}} = R^{-d + 1}\es^{-3/2
+ d/2} \propto \es^{-\t \gamma}.
\end{equation}
Equation~\eqref{sfscl4.d} predicts that 
$\t \gamma = 1$ for $d=1$, $\t \gamma = 1/2$ for $d=2$, and
$\t \gamma = 0$ for $d=3$, in contrast to the \mf\ result $\gamma=1$ for all $d$ (see Eq.~\eqref{ozsf}).
Results consistent with the predictions in
Eq.~\eqref{sfscl4.d} were found
in
$d = 1$--3 for the potential in Eq.~\eqref{kacf12}~\cite{kletal}.

If we could do a measurement on a time scale of the order of the
fluctuation lifetime $\es^{-1}$, we would see a smeared out density
fluctuation that was radially symmetric and varied periodically in the
radial direction. We expect that the dominant periodicity would be
characterized by the wave vector $k_{0}$ and that the divergent
contribution to the structure function near
the spinodal could be approximated for $k \approx k_{0}$ by
\begin{equation}
\label{sfaver1}
S_{\rm f,\,s}(k) \sim {\es^{2}\over R^{d}\es^{3 - d/2}}\!\int\!dx\,x^{d-1}
e^{i(k - k_{0})x}.
\end{equation}
Note that we used the density of a fluctuation rather than the cluster,
which is appropriate for the time averaged cluster distribution. At $k = k_{0}$ the structure function will be given by 
\begin{equation}
\label{sfaver2}
S_{\rm f,\,s}(k_0) \sim {\es^{2}\over R^{d}\es^{3 - d/2}}R^{d}\es^{- d/2} = \es^{-1},
\end{equation}
consistent with the results of
Refs.~\onlinecite{gk1} and \onlinecite{gk2}.

In summary, the structure and finite lifetime of the fundamental clusters are
responsible for the behavior of $S(k)$ near
the liquid-solid spinodal. If the peak of $S(k)$ is at
$k = 0$ as in the Ising model, the fact that measurements are made on a time scale short compared
to the fundamental cluster lifetime rather than a time scale much longer as
required by \mf\ theory makes no difference to the measured value of the exponent that
characterizes the divergence. In contrast, if the peak of $S(k)$ is at $k \neq 0$, the measurement time scale affects
the observed value of the exponent.

\section{\label{sec9} Cluster Structure and Nucleation}

In classical nucleation theory~\cite{gunt} the
nucleating droplet is assumed to be isolated, compact, and
describable as a fluctuation about a quasiequilibrium metastable state.
Classical nucleation occurs near the coexistence curve independent of the
range of interaction~\cite{uk2}. For systems with
sufficiently long-range interactions a quench near the \psp\ can lead to 
nucleation being influenced by the critical point nature of the \psp, which implies that the surface tension will vanish as the \sp\ is approached. Near the
\psp\ the surface tension will be nonzero, nucleation will occur with
a very small surface tension~\cite{ungerkl}, and the
nucleating droplet is no longer compact as in classical nucleation theory~\cite{heermannkl,ungerkl}. We refer to this form of nucleation as
{\it spinodal nucleation}.

An elegant
way to
approach nucleation theoretically was developed most fully by Langer~\cite{langer, gunt} and adapted to
spinodal nucleation by Klein and Unger~\cite{ungerkl, ku}. The Hamiltonian in Eq.~\eqref{lgwh1}
is used to calculate the free energy $F(\beta,h)$ in the equilibrium state
\begin{equation}
\label{lgwfe}
F(\beta,h) = -k_{B}T \ln\!\!\int\!\delta \phi\,e^{-\beta
H(\phi)},
\end{equation}
which can be analytically continued from the stable to the metastable
state~\cite{langer}. The nucleating droplet is associated
with the solution of the Euler-Lagrange equation obtained from the
functional derivative of the Hamiltonian~\cite{langer}:
\begin{equation}
\label{ele}
-R^{2}\nabla^{2}\phi(\xv) -2|\e|\phi(\xv) + 4\phi^{3}(\xv) -h = 0.
\end{equation}
In Eq.~\eqref{ele} $\e<0$, that is, the
temperature
$T$ is below the critical temperature. The dominant exponential part of the
nucleation probability is obtained by substituting the solution to
Eq.~\eqref{ele} into the expression for the imaginary part of the
analytically continued free energy. The details of this approach can be
found in Ref.~\onlinecite{langer} and are outlined in
Ref.~\onlinecite{gunt}.

Klein and Unger expanded the Hamiltonian in Eq.~\eqref{lgwh1} about the \mf\ spinodal as in Eq.~\eqref{lgwfepsi} and
obtained an Euler-Lagrange equation of the form~\cite{ku}
\begin{equation}
\label{eles}
-R^{2}\nabla^{2}\psi(\xv) + \lambda_{1} \dhs^{1/2}\psi(\xv) - \lambda_{2}\psi^{2}(\xv) = 0,
\end{equation}
where
$\lambda_{1}$ and
$\lambda_{ 2}$ are constants for fixed
$\e$. It is straightforward to show that the solution to Eq.~\eqref{eles} must
have the form~\cite{ku,ungerkl}
\begin{equation}
\label{ssele}
\psi(\xv) = \dhs^{1/2}f\Big({\xv\over R \dhs^{-1/4}}\Big),
\end{equation}
which implies that the difference in the order parameter density between the interior of the droplet and the
background is $\sim \dhs^{1/2}$; this difference is vanishingly small for
large
$R$ and small
$\dhs$. This small difference presents two
problems: how can we identify when and where the nucleating droplet occurs
and how can we
determine its structure. In Ising models these
problems have been solved by mapping the spinodal onto a percolation
transition and using the relation between the clusters and the
nucleating droplet.

Because the density of the nucleating droplet over the background
is the order of $\dhs^{1/2}$, the density of the \fc s is order $\Delta
h^{1/2}/R^{d} \dhs^{3/2 - d/4}$ (Eq.~\eqref{cludensp}), and the density
of a fluctuation is $\dhs^{1/2}/(\gps)^{1/2}$, the nucleating droplet is not a fundamental cluster
or a fluctuation described by the Gaussian approximation. From the
discussion in Sec.~\ref{sec7} we know that the fluctuations near a 
spinodal are generated by a random walk of the number of fundamental
clusters in the up or down direction. A possible scenario might be that the nucleating
droplet is generated by a random walk in the number of \fc s that
produces a region of density
$\dhs^{1/2}$. From the discussion in
Secs.~\ref{sec6} and \ref{sec7} such a process would require a ``walk'' of
distance $\gps$ ($[\gps]^{2}$ steps) and a time of
$\tau \sim \dhs^{-1/2}\gps$. Because the nucleation time is the
inverse of the probability~\cite{langer}, the nucleation time would
be of the order of
$\exp (\beta\gps)$.
Hence a random walk occurs too quickly to
account for the time needed to see nucleation. 

A clue to the relation between the \fc s and the nucleating droplets
is provided in Ref.~\onlinecite{monettekl} in which nucleation was
observed near the spinodal in a
$d=2$ Ising model with long-range interactions. The nucleating droplet was
identified using intervention, and it
was
found that the number of \fc s in a correlation length volume just prior to nucleation is the order of
$G_s$, implying that $G_s$ \fc s with density $\Delta
h^{1/2}/G_s^{1/2}$ coalesced into an object with a density on the
order of $\dhs^{1/2}$.

\begin{figure}[t]
\includegraphics[scale=0.75]{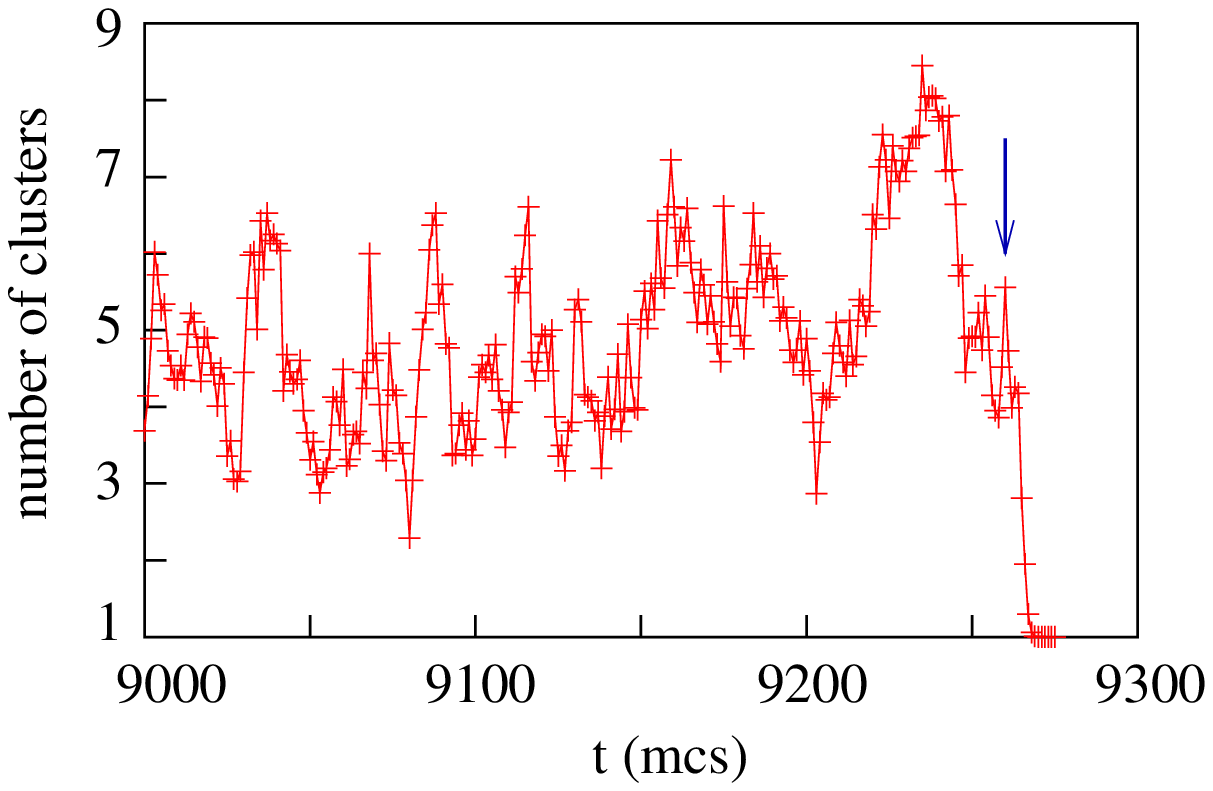}
\caption{\label{fig3}The number of \fc s in a region the size of the correlation length where nucleation will occur (see arrow). Note
the number of times where the number of \fc s is of order
$G_s$. The Monte Carlo simulation was done for a $d=2$ Ising model with $R= 28$, $h=1.25$, $G_s=12.3$, and linear dimension $L=560$.}
\end{figure}

The results of Ref.~\onlinecite{monettekl} together with the random walk argument
suggests the following picture of the relation between the
\fc s and the nucleating droplets. While the system is in the metastable
state, there are fluctuations in the number of \fc s. These fluctuations
result in regions with order $G_s=\gps$ \fc s in excess of the background on a
time scale $\sim \gps \dhs^{-1/2}$. Because this time scale is much less
than the time scale for nucleation, which is of the order of
$\exp(\beta\gps)$~\cite{ungerkl}, the appearance of 
$G_s$ clusters with a linear spatial extent of the correlation length will
happen everywhere in the system many times before the nucleation event. In
Fig.~\ref{fig3} we plot the number of \fc s the size of the
correlation length in a region where we know that nucleation will occur. Notice
the number of time intervals where the number of \fc s is of order
$G_s$. 

Because the time scale for nucleation is of the order of
$\exp(\beta\gps)$ and $G_s$ \fc s coalesce at nucleation,
there must be a free energy cost associated with the
coalescence that is the order of $G_s$. We can estimate this free energy
cost by noting that the coalescence of $G_s$ \fc s change
the density only infinitesimally. This assumption is
justified because the random walk is assumed to generate a density in the
stable phase direction of order $\dhs^{1/2}$ and the nucleating droplet has
the same density. Therefore the energy change due to coalescence is
negligible. The entropy cost can be estimated by noting that there are
$G_s$ \fc s before nucleation and one nucleating
droplet after coalescence.
These considerations imply that the entropy change is given by
\begin{equation}
\label{entropybc}
\Delta S \sim \ln 2 - \ln 2^{\gps} \sim - \gps\ln2,
\end{equation}
because each cluster has two states, up and down.
Because the energy change is negligible, the free energy change due to
coalescence is the order of 
$G_s \gg 1$. Hence the probability of nucleation and hence the probability of coalescence is $\propto \exp(-\beta\Delta F) \sim
\exp(-\beta G_s)$, in agreement with 
Refs.~\onlinecite{ungerkl} and \onlinecite{ku}.

This discussion and the one in Sec.~\ref{sec8} suggests that the \fc s are
not only a mathematical construct, but are real physical objects whose probability is not given by the usual Boltzmann factor (see Sec.~\ref{sec3}). This suggestion will be given further credence by
the discussion in the next section.

\section{\label{sec10} Clusters and models of earthquake faults}

In this section we discuss the relevance of the cluster structure to
our understanding of models of earthquake faults. The original Burridge-Knopoff model consists of blocks connected by linear springs 
to their nearest neighbors with spring constant $k_c$~\cite{bk}. 
The blocks are also connected to a loader plate by linear springs
with spring constant
$k_L$, and rest on a surface with a nonlinear velocity-weakening stick-slip
friction force. 

A simulation is initiated by choosing the displacements of the blocks at random. While
the loader plate is fixed, we determine the stress on each block (the force due to the springs) and update its velocity and displacement according to Newton's equations of motion. We continue these updates until all blocks are stuck. A block is ``stuck'' when
its velocity is below a certain threshold and other criterion are met~\cite{carllang1, carllang2, junetal}. We then add
stress to all the blocks by moving the loader plate to bring the
block with the largest stress to failure. That is, when the stress on a block 
exceeds the static coefficient of friction, the block ``fails'' and 
begins to slip. This process insures that there
is only one block that initiates the failure sequence. An earthquake is comprised of all the blocks that fail between
plate updates.

The number $n_s$ of earthquakes with $s$ blocks exhibits a power law dependence
on $s$, 
\begin{equation}
\label{bkscal}
n_{s} \sim s^{-x},
\end{equation}
with $x\approx2$ if the blocks are connected by only nearest-neighbor springs~\cite{carllang1, carllang2,junetal}.
However, the calculation of realistic stress transfer Green's functions for
real faults~\cite{steketee, rybicki} suggests that we should consider
springs that connect further neighbor blocks. The behavior of the generalized Burridge-Knopoff model is more complicated and in the limit of long-range stress transfer and a slow decrease of the velocity-weakening friction force with increasing velocity, it has been found in simulations that $x\approx 3/2$~\cite{junetal}.

To provide more insight into the behavior of the model with long-range
stress transfer we discuss a cellular automaton (CA) version of the
Burridge-Knopoff model introduced by Rundle, Jackson, and
Brown~\cite{runjack, runbrown,ofc}, which considers blocks and springs
as in the Burridge-Knopoff model. A 
failure threshold
$\sigma_{F}$ and a residual stress $\sigma_{R}$ is specified for each block. For
simplicity, we will take the $\sigma_{F}$ and $\sigma_{R}$ to be the same for
all blocks. The stress on a block is given by
$\sigma_j = k_L(\Delta - u_j) + k_c \sum_i{(u_i-u_j)}$, where $\Delta$ is the displacement of the loader plate and $u_j$ is the displacement of block $j$ from its initial position.
If
$\sigma_{j}<\sigma_{F}$, we do nothing and proceed to the next block. If 
$\sigma_{j}\geq \sigma_{F}$, we move the block a distance $\Delta u$
where
\begin{equation} 
\label{deltax}
\Delta u_{j} = {\sigma_{j} - \sigma_{R}\over k_{L} + qk_{c}},
\end{equation}
and $q$ is the number of sites within the interaction range. If the range of stress transfer is
$R$, then $q = (2R + 1)^{2}$. Once the system is quiescent, that is,
$\sigma_{j}<\sigma_{F}$ for all $j$, the plate is updated as
in the Burridge-Knopoff model.

The number of earthquakes with $s$ failed
blocks has a power law dependence as in Eq.~\eqref{bkscal} with $x = 3/2$ in the \mf\ limit $R \to \infty$. How does this scaling arise? A clue is that the long-range stress transfer CA models with noise
added to the stress drop can be described by equilibrium statistical
mechanics~\cite{runetal1, fergetal1}. It also has been shown that this
model can be described by a Langevin or Landau-Ginzburg equation in the
limit
$R\rightarrow\infty$~\cite{kleinetallg}. The latter equation has the same form as
Eq.~\eqref{lgwfepsi} (see Ref.~\onlinecite{kleinetal2}). These considerations
imply that the scaling behavior $n_s \sim s^{-3/2}$ in the long-range CA models is identical for scaling purposes to scaling
near the Ising
spinodal~\cite{kleinetal2}.

To obtain the scaling exponent $x$, we use the fact that the Ising spinodal can be
described by a Fisher droplet model~\cite{fisher} in
which the system near a critical point can be described by a collection of
non-interacting droplets. Fisher assumed that the distribution of the droplets scale as~\cite{fisher}
\begin{equation}
\label{fisherd}
{\tilde n}_{s} \sim {e^{-\dhs s^{\t \sigma}}\over
s^{\tau}}.
\end{equation}
From the cluster mapping near the spinodal discussed in
Appendix~\ref{sec5}, we know that there exists independent objects, the fundamental clusters, which scale in a 
similar way. To calculate the exponent $\tau$ for the 
earthquake CA model, we need only calculate the cluster scaling exponent for the 
Ising spinodal.

To obtain the exponent $\tau$, which we will relate to the
exponent
$x$ in Eq.~\eqref{bkscal}, we note that the Fisher droplet model exponents
are related to the spinodal exponents through the first several
moments of Eq.~\eqref{fisherd}. In particular, the isothermal susceptibility
$\chit$ is given by the second moment of ${\tilde n}_{s}$
\begin{equation}
\label{fdexp1}
\chit \propto \!\int\!ds\,s^{2}{e^{-\dhs s^{\tilde \sigma}}\over
s^{\tau}} \propto \dhs^{-1/2},
\end{equation}
and the order parameter density $\psi$ (see Appendix~\ref{sec5}) is
related to the first moment of ${\tilde n}_{s}$:
\begin{equation} 
\label{fdexp2}
\psi \propto \!\int\!ds\,s{e^{-\dhs s^{\t \sigma}}\over
s^{\tau}} \propto \dhs^{1/2}.
\end{equation}
Here we have kept $G_s$ constant and used the fact that hyperscaling holds. This constraint is appropriate for the scaling events we are considering because these models self-organize to run at a fixed distance from the spinodal with $G_s \sim 3$--5~\cite{kleinetal2}.
We can assume that the exponential in the integrals is approximately one until $s^{\t \sigma}\sim \dhs^{-1}$. Hence,
\begin{align}
\label{fdexp3}
{3 - \tau\over \t \sigma} & = {1\over 2}, \\
\noalign{\noindent and}
\label{fdexp4}
{\tau - 2\over \t \sigma} & = {1\over 2}.
\end{align}
Equations~\eqref{fdexp3} and \eqref{fdexp4} yield $\tau = 5/2$ which
apparently differs from the measured value of $x=3/2$. However, the exponents
obtained by this reasoning assume that the cluster distribution is obtained by tossing
bonds between occupied sites. In the
earthquake case the clusters are grown from a seed, the
site that is brought to failure by a loader plate update. Because for a cluster of size $s$ there are $s$
places that could have been the seed, the number of such clusters is
$sn_{s}$, where
$n_{s}$ is given in Eq.~\eqref{bkscal}~\cite{ahrostauff}. Hence, $\tau = x+1$ and hence $x = 3/2$, in agreement with the simulations of models with long-range stress transfer. The scaling form for $n_s$ with $x \approx 2$ for the usual Burridge-Knopoff model with short-range interactions is not understood.

There are other properties of the distribution of earthquakes in the
Rundle-Jackson-Brown CA model that can be obtained from consideration of the
clusters in Ising models near spinodals. We refer the interested reader to
Ref.~\onlinecite{kleinetal2}. The relation between the CA and Burridge-Knopoff
models for long-range stress transfer is discussed in Refs.~\onlinecite{xia1} and \onlinecite{xia2}.

\section{\label{sec11}Summary and conclusions}

We have shown that the structure of the clusters near
mean-field critical points and (pseudo)spinodals and its relation to thermal fluctuations is more complicated
than the corresponding relation in systems that are not mean-field and which obey hyperscaling. 
Moreover, these fundamental clusters and their structure have physical consequences which implies that the 
clusters are not only convenient mathematical constructs. 

For \nnmf\ systems the thermal fluctuations are generated by fluctuations in the number
of \fc s. Because the clusters in \nnmf\ systems play a different role and appear to be real 
physical objects rather than just mathematical constructs, we refer to them as \fc s. The probability of finding a \fc\ is not given by the Boltzmann factor because the cluster lifetime 
is much less than the decorrelation time. Their physical consequences
are seen most clearly near the \psp\ in
supercooled liquids where the relation between the measurement time and
the lifetime of the \fc s yields predictions for the behavior of the
structure function that are confirmed by simulations on \nmf\ systems and are contrary to \mf\
theory.

In addition to the applications of the cluster structure we have discussed, there are many other applications that have shed light on physical
processes. These applications include:

\begin{enumerate}

\item The elucidation of the early time structure of systems
undergoing spinodal decomposition and continuous ordering, including the
understanding of why the linear theory of Cahn, Hilliard, and
Cook~\cite{cahn,hilliard, cook} fails first at large momentum
transfer~\cite{grossetal1, grossetal2}, the fractal structure of the mass
distribution of early time spinodal decomposition~\cite{klein1}, and a
physical interpretation of the fermionic (Grassman) variables associated
with a supersymmetric representation of the early stage continuous
ordering~\cite{grossetal1, grossetal2, kleinbat}.

\item The phase separation of polymer and solvent in the presence of
gelation~\cite{conigliostakl}.

\item Possible precursors to nucleation near the
\psp~\cite{monettekl}.

\end{enumerate}

Future work includes the application of cluster methods to the study of
precursors to large earthquakes in the CA and Burridge-Knopoff models, the
study of heterogeneous nucleation near \psp s, the investigation of fracture
and the merging of microcracks, and the study of the crossover from the
linear regime of spinodal decomposition and continuous ordering to
nonlinear evolution.

\appendix

\section{\label{sec5} Percolation Mapping}

To obtain a deeper understanding of the structure of the
fluctuations near the \mf\ critical point, we map
the Ising critical point (\mf\ and non-mean-field) onto a
percolation transition for a properly chosen percolation model~\cite{ck}.
We first describe the mapping introduced by Kasteleyn and Fortuin~\cite{kf}
of the $s$-state Potts model onto random bond percolation. The latter is
defined on a lattice where all the sites or vertices are
occupied with probability one, and the bonds are occupied with a
probability $p_{b}$. Clusters are defined as a set of sites connected to
each other by bonds and not connected to any other sites in the
lattice~\cite{ahrostauff}.

The Hamiltonian for the $s$-state Potts model is
\begin{equation}
\label{potts1}
H_{\rm P} = -\jp \sum_{i,j}(\delta_{\sigma_{i}\sigma_{j}} - 1)
-\hp (\delta_{\sigma_{i}1} - 1),
\end{equation}
where $\sigma_{i}$ specifies the state of site
$i$,
$\jp>0$ is the coupling constant, and $\hp$ is the Potts field;
the Kronecker delta $\delta_{\sigma_{i}\sigma_{j}}\neq 0$ only
when sites $i$ and $j$ within the interaction range are in
the same state. We first set $\hp = 0$ and
write the Boltzmann factor
$e^{-\beta H_P}$ as
\begin{equation}
\label{potts3}
e^{-\beta H_{\rm P}} = \prod_{ij}\big[
\delta_{\sigma_{i}\sigma_{j}} + e^{-\beta \jp}(1 -
\delta_{\sigma_{i}\sigma_{j}})\big], 
= \prod_{ij} \big[ (1 -
e^{-\beta \jp})\delta_{\sigma_{i}\sigma_{j}} + e^{-\beta \jp}\big].
\end{equation}
We associate a bond with $\delta_{\sigma_{i}\sigma_{j}} = 1$ and the
absence of a bond with $\delta_{\sigma_{i}\sigma_{j}} = 0$. With this
association the generating
function for the random bond percolation model is obtained by
differentiating the free energy for the $s$-state Potts model with respect
to $s$ and then setting $s$ equal to 1~\cite{kf}. There are many subtle
mathematical points in the Kasteleyn-Fortuin proof of this relation, and we
refer the reader to Ref.~\onlinecite{kf} and the references 
therein for the details. Because it will be needed to understand the
structure of fluctuations, we demonstrate how the connection between the
Potts model and percolation works. 

The partition function $Z_{\rm P}$ is given by
\begin{equation}
\label{potts4}
Z_{\rm P} = \sum_{\sigma} e^{-\beta H_{\rm P}}.
\end{equation}
and the free energy in the canonical ensemble is 
$F_{\rm P}(\beta,s) = -k_{B}T\ln Z_{\rm P}$.
If we differentiate $-\beta F_{\rm P}(\beta,s)$ with
respect to $s$, we obtain 
\begin{equation}
\label{potts6}
-\beta{\pa F_{\rm P}(\beta,s)\over \pa s} = \sum_{\sigma}{1 \over Z_{\rm P}}
{\pa\,e^{-\beta H_{\rm P}}\over \pa s}.
\end{equation}
Setting $s = 1$ results in $H_{\rm P} = 0$ because there is only one Potts
state; hence $Z_{\rm P} = s^{N} = 1$ for $s = 1$, where $N$ is the
number of sites in the lattice. Therefore
\begin{equation}
\label{potts7}
-\beta{\pa F_{\rm P}(\beta,s)\over \pa s}\Big|_{s=1} = \frac{\pa} {\pa s}\sum_{\sigma}
e^{-\beta H_{\rm P}} \Big|_{s=1}.
\end{equation} 

The percolation generating function is the right-hand side of
Eq.~\eqref{potts7}. To understand this interpretation we consider
several terms in Eq.~\eqref{potts7}. We use Eq.~\eqref{potts3} for
$e^{-\beta H_{\rm P}}$ and first consider the term
$e^{-\beta \jp}$ in each of the factors in 
the product; that is, we include no terms with 
$\delta_{\sigma_{i}\sigma_{j}}$. For a lattice with $c=qN/2$ total possible bonds, we find a contribution to 
$F_{\rm P}(\beta, s)$ of the form $s^{N}e^{-\beta \jp cN}$. By
differentiating with respect to $s$ and setting $s=1$, we obtain the
contribution to the generating function $G_{\rm f}$:
\begin{equation}
\label{gen1} 
G_{{\rm f}, 1} = Ne^{-\beta \jp cN}.
\end{equation}
Because there are $N$ sites and $cN$ bonds, $G_{\rm f,\,1}$ can
be interpreted as the mean number of single site clusters. That is, $e^{-\beta \jp cN}$ is the probability that there
are no bonds present. 

We now consider a term from Eq.~\eqref{potts3} that includes only one delta
function, which we take to be
$\delta_{\sigma_{1}\sigma_{2}}$. The contribution to
$G_{\rm f}$ has the form
\begin{equation}
\label{gen2}
G_{\rm f,\,2}(p) = (1 - e^{-\beta \jp})se^{-\beta \jp(cN-1)}s^{N-2} =
p_{b}(1 - p_{b})^{cN-1}s^{N-1},
\end{equation}
where we have associated the bond probability $p_{b}$ with $1 -
e^{-\beta \jp}$. Differentiating with respect to $s$ and setting $s = 1$ 
gives $N-1$ for the number of clusters times the probability of such a
configuration. There are $N - 2$ one-site clusters and one two-site
cluster for a particular bond. The number of ways we
can choose one bond is $cN$, so that the first two contributions to 
$\pa F(\beta, s)/\pa s$ are
\begin{equation}
\label{gen3}
G_{\rm f,\,1}(p) + G_{\rm f,\,2}(p) = N(1-p_{b})^{cN} + (N -
1)cNp_{b}(1 - p_{b})^{cN - 1}.
\end{equation} 
If we continue in this manner, we would find that the terms we
obtain are the number of clusters in a given configuration. The complete
enumeration of the configurations will lead to an expression for the mean
number of clusters as a function of $p_{b}$. 
Hence, the mean number of clusters can be written as 
\begin{equation}
\label{gen4}
G_{\rm f}(p) = \sum_{k} \lb n_{k}\rb,
\end{equation}
where $\lb n_{k}\rb$ is the mean number of clusters with $k$ sites.

To obtain the full generating function for random bond percolation, we
must include the field $\hp$ in the calculation of the free energy.
We write
\begin{equation}
\label{gen5}
e^{-\beta H_{\rm P}} = \prod_{ij}\big[ (1 -
e^{-\beta \jp})\delta_{\sigma_{i}\sigma_{j}} + e^{-\beta \jp}\big]
\prod_{l}\big[ (1 - e^{-\beta\hp})
\delta_{\sigma_{l}1} + e^{-\beta\hp}\big].
\end{equation} 
The terms in the expansion of Eq.~\eqref{gen5} represent sets of
connected sites (clusters) generated by
$\delta_{\sigma_{i}\sigma_{j}}$. Terms of the form
$\delta_{\sigma_{k}1}$, where the index $1$ labels one of the $s$ possible
states of a site, give no contribution to
the derivative of the partition function with respect to $s$ because the
$\delta_{\sigma_{k}1}$ term fixes all spins in a cluster to the Potts state
labeled as 1. Hence there is no $s$ dependence and no factor of $s$ in the 
product, which equals $s$ raised to the power of the number of clusters. From Eq.~\eqref{gen5} all clusters with a nonzero
weight after differentiation with respect to $s$ will have a field dependence
of the form $e^{-n\hp}$, where $n$ is the number of sites in the cluster. If
we resum as in Eq.~\eqref{gen3}, we obtain~\cite{kf} 
\begin{equation}
\label{gen6}
G_{\rm f}(p_b,\hp) = \sum_{k} \lb n_{k}\rb\,e^{-k \beta \hp}.
\end{equation}
The
reader might want to work out the generating function for small lattices to
see how the sum in Eq.~\eqref{gen6} arises.

To investigate the mapping of the percolation model onto the Ising model,
we consider the dilute $s$-state Potts~\cite{mur} model with the
Hamiltonian 
\begin{equation}
\label{dpgen1}
\beta H_{\rm DP} = -\beta\jp\sum_{ij} (\delta_{\sigma_{i}\sigma_{j}} - 1)n_{i}n_{j}
- \beta\hp \sum_{i}(\delta_{\sigma_{i}1} - 1)n_{i} - K_{\rm LG}\sum_{ij}
n_{i}n_{j} +
\Delta\sum_{i} n_{i}.
\end{equation}
where
$n_{i} = 1$ denotes that a particle (spin) occupies site $i$ and
$n_{i} = 0$ denotes an empty site. Hence there is a Potts interaction between
occupied ($n_{i} = 1$) sites. The quantity 
\begin{equation}
\label{dpgen2}
\beta H_{\rm LG} = -K_{\rm LG}\sum_{ij} n_{i}n_{j} + \Delta \sum_{i}n_{i}
\end{equation}
is the Hamiltonian for the lattice gas formulation of the Ising
model~\cite{huang}, with $K_{\rm LG}$ the (dimensionless) coupling
constant and $\Delta$ the chemical potential. In terms of the parameters
in the Ising Hamiltonian
$H_{\rm I}$,
\begin{equation}
\label{HI}
\beta H_{\rm I} = -K_{\rm I}\sum_{ij}s_{i}s_{j} + \beta h_{\rm I}\sum_{i}s_{i}
\end{equation}
with $s_{i} = \pm 1$, we have~\cite{huang}
\begin{equation}
\label{lgi}
K_{\rm LG} = 4K_{\rm I}\ \mbox{and}\ \Delta = \beta h_{\rm I} + 2cK_{\rm I}.
\end{equation} 
Differentiating the free energy constructed from
$H_{\rm DP}$ with respect to $s$ and setting $s = 1$ results in the
generating function for correlated site random bond percolation for which
occupied sites are distributed according to the lattice gas
Hamiltonian in Eq.~\eqref{dpgen2} and bonds are thrown randomly with a probability
$p_{b}$ between pairs of occupied sites. To understand this connection we
note that 
\begin{equation}
\label{zpotts}
Z_{\rm DP} = \sum_{\sigma_{i}\sigma_{j}n_{i}n_{j}} e^{-\beta
H_{\rm DP}},
\end{equation}
and
\begin{equation}
\label{dpfree}
-{\pa k_{B}T\ln Z_{\rm DP} \over \pa
s}\Big |_{s=1} = {{\pa \over \pa s}(
\sum_{\{\sigma_{i}\} \{ n_{i}\}} e^{
\sum_{ij} \beta \jp(\delta_{\sigma_{i}\sigma_{j}} - 1)n_{i}n_{j} +
\beta \hp \sum_{i}(\delta_{\sigma_{i}1} - 1) 
n_{i}}e^{-\beta H_{\rm DP}})\Big|_{s=1}  \over Z_{\rm
DP}}.
\end{equation}
If we write
\begin{subequations}
\begin{align}
\label{DPmap1}
\exp\big[\beta \jp(\delta_{\sigma_{i}\sigma_{j}} -
1)n_{i}n_{j}\big] & = \lbrack (1 -
e^{-\beta \jp})\delta_{\sigma_{i}\sigma_{j}} + e^{-\beta \jp}\rbrack n_{i}n_{j} +
(1 - n_{i}n_{j}), \\
\noalign{\noindent and}
\label{DPmap2}
\exp\big[ \beta \hp \sum_{i}(\delta_{\sigma_{i}1} -
1)n_{i}\big] & = (1 - e^{- \beta \hp })\delta_{\sigma_{i}1}n_{i} 
+ e^{-\beta \hp}n_{i} +
(1 - n_{i}),
\end{align}
\end{subequations}
we can use the same arguments that we gave for random percolation to show
that the expression in Eq.~\eqref{dpfree} leads to the generating function
for correlated site random bond percolation where the occupied sites are
distributed according to the lattice gas Boltzmann factor constructed from
the Hamiltonian in Eq.~\eqref{dpgen2}~\cite{mur}. Note that for this model
the sum over the $s$ states of the Potts spin is one for all $s$ if site $i$
is empty. 

We now consider the Hamiltonian $H_{\rm DP}$ in Eq.~\eqref{dpgen1}. We set
$\hp=h_{I}=0$, and hence 
$\Delta = 2qK_{\rm I} = qK_{\rm LG}/2$ from Eq.~\eqref{lgi} and let 
$\jp = K_{\rm LG}/2$ in Eq.~\eqref{dpgen1}. Then $H_{\rm DP}$ becomes
\begin{equation}
\label{hdp1}
\beta H_{\rm DP} = -{K_{\rm LG}\over
2}\sum_{ij}(\delta_{\sigma_{i}\sigma_{j}} - 1)n_{i}n_{j} -
K_{\rm LG}\sum_{ij}n_{i}n_{j} + {K_{\rm LG}\over 2}\sum_{ij}( n_{i} +
n_{j}).
\end{equation}
Suppose that for a pair of sites within the interaction range, either both
sites are empty or both sites are filled, but the spins are in the same
Potts state. When both sites are empty, there is only one Potts
configuration. When both sites are occupied and in the same Potts state, there are $s$ configurations, Hence, there are
$s + 1$ configurations with $H_{\rm DP} = 0$. Now
consider the case where either one site is occupied and one empty or
both sites are occupied, but the spins are in different Potts states. In
this case the contribution of this pair to $H_{\rm DP}$ is $K_{\rm LG}/2$,
and the number of ways this combination can be obtained is $(s + 1)s$. These
considerations imply that the dilute $s$-state Potts model at
$\jp = K_{\rm LG}/2$ is equivalent to a pure $(s + 1)$-state Potts model
with the Hamiltonian
\begin{equation}
\label{ppotts}
\beta H_{{\rm P},\,(s+1)} = -{K_{\rm LG}\over
2}(\delta_{\sigma_{i}\sigma_{j}} - 1),
\end{equation}
where $\sigma_{i}$ can be in $s + 1$ states. The $s = 2$
Potts model 
is the lattice gas model in Eq.~\eqref{dpgen1} with $\Delta =
cK_{\rm LG}/2$. That is, for $\beta \jp = K_{\rm LG}/2 = 2K_{\rm I}$ and $h_{I}
= \hp = 0$, the Hamiltonian of the dilute $s$-state 
Potts model is the same as the Hamiltonian of the $(s + 1)$-state pure
Potts model. In the limit $s\rightarrow 1$, the $(s + 1)$-state pure
Potts model is the lattice gas model. 

In this formulation we can write the Boltzmann factor as 
\begin{align}
\label{pottsbf}
e^{-\beta H_{{\rm P},\,(s+1)}} & = \prod_{ij}[ (1 -
e^{-{ K_{\rm LG}/2}})\delta_{\sigma_{i}\sigma_{j}} + e^{-{ K_{\rm
LG}/2}}] \\
\label{pottsbf2}
& = e^{-{K_{\rm LG}cN/
2}}\prod_{ij}\Big\lbrack {1 - e^{-{K_{\rm LG}/2}}\over
e^{-{K_{\rm LG}/2}}}\delta_{\sigma_{i}\sigma_{j}} + 1\Big\rbrack.
\end{align}
Clearly the singular behavior of the free energy comes from the terms in Eq.~\eqref{pottsbf2} contained in
the product over lattice sites. This product has the form of
$\prod_{ij}\lbrack f_{ij} +1\rbrack$, where we associate a graph or
cluster with a product of the
$f_{ij}$ summed over Potts states. These clusters are the same as the
percolation clusters because the sites are connected by
$\delta_{\sigma_{i}\sigma_{j}}$ bonds. The linked cluster
theorem~\cite{uf,ng} states that the singular part of the free energy
$F_{\rm P,\,sing}$ is the sum over all {\it connected} graphs
in the thermodynamic limit. Connected graphs are those in which all points or
vertices of the graph are connected by an $f_{ij}$ bond.
Because the graphs are connected, the sum over Potts states results in a
factor of
$s$, independent of the size or structure of the graph. Therefore the
derivative of $F_{\rm P,\,sing}$ with respect to $s$ results in the same sum
without the overall factor of $s$. Thus
\begin{equation}
\label{link}
2 {dF_{\rm P,\,sing}\over ds}\big|_{s=1} = F_{\rm P,\,sing}\big |_{s=1},
\end{equation}
which implies that the percolation transition and the Ising critical point occur at the same temperature and have the same
critical exponents. The amplitudes of the singular quantities differ by a
factor of two. If instead of the Hamiltonian in Eq.~\eqref{dpgen1},
which defines clusters as consisting of only occupied sites, we add a term
to the Hamiltonian of the form
\begin{eqnarray}
\label{sym}
\beta H_{\rm DP,\,empty} &=& -\beta \jp\sum_{ij}(\delta_{\sigma_{i}\sigma_{j}} -
1)(1 - n_{i})(1 - n_{j}) \nonumber \\
&&{}-K_{\rm LG}\sum_{ij}(1 - n_{i})(1 - n_{j}) +
\Delta^{\prime}\sum_{i}(1 - n_{i}),
\end{eqnarray}
we can define clusters between ``empty'' sites. In this way
the singular part of the free energy will be identical to the mean number of
clusters for
$s\rightarrow 1$ and $\hp = 0$. The same result was obtained using
renormalization group techniques in Ref.~\onlinecite{ck}. 
The mapping of the Ising model onto a properly chosen percolation 
model is not restricted to nearest neighbor interactions. If the Ising
interaction has a range $R$, we need only to choose the Potts interaction
to also have a range $R$, which means that the bond between sites is
randomly placed between any two spins within the interaction range.

We now consider the mapping of a thermal problem onto a 
percolation model near the
spinodal. This mapping will require a slightly different approach. 
We again begin with the dilute $s$-state Potts
model. The Hamiltonian is the sum of $H_{\rm DP}$ and $H_{\rm LG}$ 
from Eqs.~\eqref{dpgen1} and \eqref{dpgen2}. This Hamiltonian can be put
into a continuum form using the Gaussian transformation~\cite{Amit,cl}.
Because we are interested in the
\mf\ limit, the
\lgw\ Hamiltonian is the free energy. We have from
Refs.~\onlinecite{klein1} and \onlinecite{cl}
\begin{align}
\label{feftdp}
F_{\rm DP}(\zeta,\phi) =&\!\!\int\! d\xv \Big[ {1\over 2}
s(s-1)\big( R\nabla\zeta(\xv)\big)^{2} - r_{1}s(s -
1)\zeta^{2}(\xv) - \hp(s-1)\zeta(\xv) \nonumber \\
&+ {w_{1}\over 4!}s(s - 1)(s - 2)(s -
3)\zeta^{3}(\xv) + {w_{2}\over 2}s(s - 1)\zeta^{2}(\xv)\phix\!\Big] + F(\p),
\end{align} 
where $F(\p)$ is the free energy in Eq.~\eqref{lgfe}. 
The constants
$r_{1},\, w_{1},\, w_{2}$, and $\epsilon$ in Eq.~\eqref{lgfe} can be written
as functions of
$J,\, K_{\rm LG}$, and $c$. The global
percolation order parameter $\zeta$ is the probability that a spin in the
stable phase direction belongs to the infinite cluster of occupied sites. As
for a discrete system, the percolation model is obtained by
differentiating $F_{\rm DP}$ with respect to $s$ and setting
$s = 1$~\cite{cl}.

To map the thermal problem near the spinodal onto the
percolation problem, we rewrite Eq.~\eqref{lgfe} as in Eq.~\eqref{lgwfepsi}
with
$\dhs = h_{s} - h_I$. We equate the functional derivative of $dF_{\rm
DP}(\zeta, \phi)/ds|_{s=1} = F_{\rm P}$ with respect to $\zeta(\xv)$ to
the functional derivative of
$F(\orders)$ with respect to $\orders$. That is,
\begin{align}
\label{ftpotts}
{\delta F_{\rm P}\over \delta \zeta} &= -R^{2}\nabla^{2}\zeta(\xv) -
2r_{1}\zeta(\xv) + {w_{1}\over 3}\zeta^{2}(\xv) + w_{2}\zeta(\xv)\p(\xv)
- \hp, \\
\noalign{\noindent and}
\label{ftising}
{\delta F(\orders)\over \delta \orders} &= -R^{2}\nabla^{2}\orders(\xv) +
2\lambda_{1} \dhs ^{1/2}\orders(\xv) - 3\lambda_{2}\orders^{2}(\xv) +
4\lambda_{3}\orders^{3}(\xv) 
\end{align} 
must be equal. This condition implies that
$
-2(r_{1} - {w_{2}\over 2}\phi_{s}) \zeta + {w_{1}\over 3}\zeta^{2} - \hp
$
must be the same as 
$
2\lambda_{1} \dhs^{1/2}\orders -3\lambda_{2}\orders^{2}$.
We dropped the $\orders^{3}$ term in the Euler-Lagrange equation 
because $\orders\sim \dhs^{1/2}\ll 1$ as we shown in Secs.~VI and
XI.

At the spinodal ($\dhs = 0$) we identify $\zeta$ with $-\orders$
and require that $\hp = 0$, $w_{1}/3 = -3\lambda_{2}$, and $r_{1} =
w_{2}\p(\xv)/2$. With these equalities the solutions of
Eqs.~\eqref{ftpotts} and \eqref{ftising} with $\delta F_P/\delta
\zeta(\xv) = \delta F(\orders)/
\delta
\orders(\xv) = 0$ are identical. If we write the
parameters in Eqs.~\eqref{ftpotts} and \eqref{ftising} in terms of the parameters $\jp,\, K_{\rm LG}$, and $c$ of $H_{\rm DP}$ in
Eqs.~\eqref{dpgen1} and
\eqref{dpgen2}, we obtain 
\begin{equation}
\label{map}
\beta\jp = 2K_{I}(1- \rho),
\end{equation}
where the density $\rho = (1 + m)/2$ and $\p$ is proportional to $m$, the
magnetization per spin. Because we are interested in the coincidence of the 
spinodal with a percolation transition, we can set $\p(\xv) = \p_{s}$,
the value of the order parameter at the spinodal.

To define the bond probability in terms of a physically meaningful
quantity, we need to obtain the proportionality factor between $\p_{s}$ and
$m_{s}$, the value of the magnetization at the spinodal. To do so we use the
relation between the Ising model and the $\phi^{4}$ theory
generated by the Hubbard-Stratonovich transformation~\cite{Amit}. In
particular, the value of
$\p_s$ is~\cite{cl}
\begin{equation}
\label{phispin}
\p_{s} = \pm{ [(cK_I - 1)c]^{1/2}\over c^{2}K_I}.
\end{equation}
When $T\rightarrow 0$ or equivalently $K_I\rightarrow \infty$, 
$m_{s} \to 1$ as can be seen by noting that the magnetic
field is divided by $T$ in the Boltzmann factor. Because
$\p_{s}\rightarrow 1/cK_I^{1/2}$ as
$K_I\rightarrow \infty$, we have
\begin{equation}
\label{phim}
m_{s} = cK_I^{1/2}\p_{s},
\end{equation}
so that $m_s \to 1$ as $K_I\rightarrow \infty$. If we use Eq.~\eqref{phim}, we
obtain the expression for the bond probability that maps the percolation
model onto the spinodal 
\begin{equation}
\label{bp}
p_{b} = 1 - e^{-\beta \jp},
\end{equation}
where $\jp$ is given in Eq.~\eqref{map}. The
validity of this mapping was demonstrated numerically in
Refs.~\onlinecite{heermannkl,monettekl}, and \onlinecite{raykl}.

The interpretation of this mapping is that the spinodal curve is the locus
of a set of percolation transitions. If the spinodal curve is approached
with $h \neq 0$, there is a transition to a spanning cluster in
the stable phase direction at the spinodal. We stress that this result is
correct only in the 
\mf\ limit, $G_{s}\rightarrow \infty$. 

The various mappings we have discussed imply that the free energy of the
lattice gas model is isomorphic to the generating function for correlated
site random bond percolation. For $\hp=h_I=0$, the generating function is
the mean number of clusters.

\begin{acknowledgements}

We are pleased to acknowledge useful conversations with M.\ Anghel, G.\ G.\
Batrouni, A.\ Coniglio, C.\ Ferguson, G.\ Johnson, L.\ Monette, T.\ Ray, and
P.\ Tamayo. Klein and Rundle received support from the DOE grants 
DE-FG02-95ER14498 and DE-FG03-95ER14499, respectively, Tiampo was supported by an 
NSERC Discovery Grant, and Gould and Gulbahce were supported in 
part by NSF DUE-0127363. N. Gulbahce was also supported by Center for Nonlinear 
Studies and LDRD program at LANL. This work was partly carried out under the auspices of 
the National Nuclear Security Administration of the U.S.\ Department of Energy at 
Los Alamos National Laboratory under Contract No.\ DE-AC52-06NA25396.

\end{acknowledgements}

\end{document}